\documentclass[aps,prc,reprint,showpacs,groupedaddress,onecolumn]{revtex4-1}

\usepackage{graphicx,color} 
\usepackage{dcolumn}  
\usepackage{bm}       
\usepackage{amsmath}
\usepackage[dvipsnames]{xcolor}

\begin{document}

\newcommand {\nc} {\newcommand}

\newcommand{\vv}[1]{{$\bf {#1}$}}
\newcommand{\ul}[1]{\underline{#1}}
\newcommand{\vvm}[1]{{\bf {#1}}}
\def\btau{\mbox{\boldmath$\tau$}}

\nc {\IR} [1]{\textcolor{red}{#1}}
\nc {\IB} [1]{\textcolor{blue}{#1}}
\nc {\IP} [1]{\textcolor{magenta}{#1}}
\nc {\IM} [1]{\textcolor{Bittersweet}{#1}}
\nc {\IE} [1]{\textcolor{Plum}{#1}}

\nc{\ninej}[9]{\left\{\begin{array}{ccc} #1 & #2 & #3 \\ #4 & #5 & #6 \\ #7 & #8 & #9 \\ \end{array}\right\}}
\nc{\sixj}[6]{\left\{\begin{array}{ccc} #1 & #2 & #3 \\ #4 & #5 & #6 \\ \end{array}\right\}}
\nc{\threej}[6]{ \left( \begin{array}{ccc} #1 & #2 & #3 \\ #4 & #5 & #6 \\ \end{array} \right) }
\nc{\half}{\frac{1}{2}}
\nc{\numberthis}{\addtocounter{equation}{1}\tag{\theequation}}
\nc{\lla}{\left\langle}
\nc{\rra}{\right\rangle}
\nc{\lrme}{\left|\left|}
\nc{\rrme}{\right|\right|}

\title{\textit{Ab initio}  Translationally Invariant Nonlocal One-body Densities\\ from  No-core Shell-model Theory} 

\author{M. Burrows$^{(a)}$}
\author{Ch. Elster$^{(a)}$}
\author{G.~Popa$^{(a)}$}
\author{K.D. Launey$^{(b)}$}
\author{A. Nogga$^{(c)}$}
\author{P.~Maris$^{(d)}$}

\affiliation{(a)Institute of Nuclear and Particle Physics,  and
Department of Physics and Astronomy,  Ohio University, Athens, OH 45701,
USA \\
(b) Department of Physics and Astronomy, Louisiana State University,
Baton Rouge, LA 70803, USA\\
(c) IAS-4, IKP-3, JHCP, and JARA-HPC,  Forschungszentrum J\"ulich, D-52428
J\"ulich, GER \\
(d) Department of Physics and Astronomy, Iowa State University, Ames, IA 50011, USA
}

\date{\today}

\begin{abstract}
\begin{description}
\item[Background] It is well known that effective nuclear interactions are in general nonlocal. Thus if
nuclear densities obtained from {\it ab initio} no-core-shell-model (NCSM) calculations are to be used in
reaction calculations,  translationally  invariant nonlocal densities must be available.  

\item[Purpose] Though it is standard to extract translationally invariant one-body local densities from
NCSM calculations to calculate local nuclear observables like radii and transition
amplitudes, the corresponding nonlocal one-body densities have not been considered so far. A major
reason for this is that the procedure for removing the center-of-mass component from NCSM
wavefunctions up to now has only been developed for local densities.

\item[Results] A formulation for removing center-of-mass contributions from nonlocal one-body
densities obtained from NCSM and symmetry-adapted NCSM (SA-NCSM) calculations is derived, and applied to the ground state densities of
$^4$He, $^6$Li, $^{12}$C, and $^{16}$O. 
The nonlocality is studied as a function of angular momentum components in
momentum as well as coordinate space. 

\item[Conclusions]  We find that the nonlocality for the ground state densities of the nuclei under
consideration increases as a function of the angular momentum. The relative magnitude of those
contributions decreases with increasing angular momentum. In general, the nonlocal structure of the
one-body density matrices we studied is given by the shell structure of
the nucleus, and can not be described with simple
functional forms.

\end{description}
\end{abstract}

\pacs{21.60De,27.20.+n}

\maketitle

\section{Introduction and Motivation}
\label{intro}

Recent developments of the nucleon-nucleon ($NN$) or three-nucleon ($3N$)
interactions, derived from chiral effective field theory, have yielded major progress
\cite{EntemM03,Epelbaum06,Epelbaum:2008ga}. 
These, coupled with the utilization of massively parallel
computing resources (e.g., 
see~\cite{LangrSTDD12,shao2016,CPE:CPE3129,Jung:2013:EFO}),
 have placed
\textit{ab initio} large-scale simulations at the frontier of nuclear structure and reaction explorations. Among other successful many-body theories,
the  {\it ab initio} no-core shell-model (NCSM) approach, which
has considerably advanced our understanding and capability of
achieving  first-principles descriptions of low-lying states in light nuclear
systems~\cite{Navratil:2000ww,Navratil:2000gs,Roth:2007sv,BarrettNV13,Stumpf:2015lma}, has over the last decade taken center stage
in the development of microscopic tools for studying the structure of atomic nuclei. The NCSM concept combined with a symmetry-adapted (SA) basis  in the {\it ab initio} SA-NCSM \cite{DytrychLMCDVL_PRL12} has further expanded the reach to the structure of intermediate-mass nuclei \cite{LauneyDD16}. The NCSM
framework has been successfully extended to reactions of light nuclei at low
energies (see
e.g.~\cite{Navratil:2010jn,Quaglioni:2008sm,Quaglioni:2009mn,Dohet-Eraly:2015fsa}) by
combining the NCSM with resonating group methods. While this approach treats the many-body
scattering problem completely microscopically, reactions involving heavier nuclei or reactions
at higher energies are usually treated by reducing the many-body degrees of freedom to a more
manageable few-body problem and thus introducing effective interactions between relevant 
clusters.  Those effective interactions may either be phenomenologically described by
fitting e.g. scattering data, or one may attempt to extract them from structure calculations
combined with the continuum. A path along this line has recently been
proposed~\cite{Rotureau:2016jpf} based on the coupled-cluster approach to nuclear
structure.  

Microscopic folding models for those effective interactions also have a long tradition.
However, their main disadvantages is that they were usually constructed for closed
shell nuclei using relatively simple models for the nuclear structure input (see
e.g.~\cite{Arellano:1990zz,Crespo:1992zz,Elster:1996xh}). In order to open the path to account for the full microscopic structure of the clusters and employ first-principle wave functions, as those derived in the {\it ab initio} NCSM, 
it is an important first step to construct a one-body density, which is both nonlocal and translationally invariant, starting from 
one-body density matrix (OBDM) elements obtained from NCSM calculations. The need for nonlocal
densities has been recognized in reaction theory, e.g., in treating the antisymmetrization between
two localized clusters that accounts for particle exchange~\cite{NavratilQSB09}, as well
as in folding calculations of microscopic optical potentials~\cite{Arellano:1990zz,Elster:1996xh}.

In this work we present a `proof-of-principle' study that focuses on obtaining
translationally invariant ($ti$)
nonlocal one-body densities and discuss their properties. 
We concentrate on the deformed oblate $^{12}$C nucleus
and  the open-shell $^6$Li. As examples for  closed shell nuclei we consider $^{4}$He and $^{16}$O. 
The NCSM calculations employed here are carried out with the $J$-matrix inverse scattering
potential, JISP16~\cite{Shirokov:2003kk,Shirokov:2004ff}. In Sec.~\ref{formal} we first define
the nonlocal density, and then show how to remove the center-of-mass (c.m.) contribution to
arrive at a  translationally invariant  nonlocal density. In Sec.~\ref{results}, we illustrate
the off-shell structure of the $ti$ nonlocal density for $^4$He, $^6$Li, $^{12}$C, and $^{16}$O 
in momentum space as well as for $^6$Li and $^{12}$C  in
coordinate space. We also investigate the dependence of the nonlocality on the model
space, and finally provide some more details of the nonlocal structure. We summarize in
Sec.~\ref{summary}.


\section{Formal Considerations}
\label{formal}

\subsection{Space-fixed Nonlocal Densities}
\label{sfixed}

\subsubsection{Space-fixed nonlocal one-body density in coordinate space}

As a starting point we first derive a space-fixed ($sf$) nonlocal one-body density, 
$\rho_{sf}(\vec{r},\vec{r'})$,
between an initial $A$-body wave function $|\Psi \rangle$ and a final $A$-body wave function 
$|\Psi' \rangle$, 
\begin{equation}
\label{1.1.1}
\rho_{sf}(\vec{r},\vec{r'}) = \lla \Psi' \left| \sum_{i=1}^A \delta^3(\vec r_i- \vec
r) \delta^3(\vec r'_i-\vec{r'}) \right| \Psi \rra~.
\end{equation}
The many-body wave function $|\Psi\rangle$ is expanded in a basis of Slater determinants of
single-particle harmonic oscillator (HO) states. Since we use $sf$ single-particle coordinates, the wave
functions and implicitly the calculated OBDM will include the c.m. that needs to be removed later. 
In this paper OBDM elements are calculated within the 
NCSM, using the JISP16 $NN$ interaction~\cite{Shirokov:2003kk,Shirokov:2004ff}. The NCSM uses a finite set of
single-particle HO states, characterized by two basis parameters, the HO energy $\hbar \omega$ and the many-body basis space cut-off
$N_{\max}$, where $N_{\max}$ is defined as the maximum number of oscillator quanta
above the valence shell for that nucleus.

Expanding the delta functions from Eq.~(\ref{1.1.1}) in terms of spherical harmonics,  labelling the
$A$-nucleon many-body eigenstates by the total angular momentum $J$, its projection
$M$, and all additional 
quantum numbers collectively by $\lambda$, we obtain
\begin{equation}
\label{1.1.2}
\hspace*{-1cm} \rho_{sf}(\vec{r},\vec{r'}) = \lla A \lambda' J' M' \left| \sum_{i=1}^A
\frac{\delta(r_i-r)}{r^2} \frac{\delta(r_i'-r')}{r'^{2}} \sum_{l m} \sum_{l' m'}
Y_{l}^{m}(\hat{r}_i) Y_{l}^{*m}(\hat{r}) Y_{l'}^{*m'}(\hat{r}') Y_{l'}^{m'}(\hat{r}'_i) \right| A \lambda J M \rra~.
\end{equation}
Here $\hat r$ represents the angular part of vector $\vec r$.
After coupling the spherical harmonics to bipolar harmonics, 
\begin{eqnarray}
\label{1.1.3}
\mathcal{Y}^{l_1 l_2}_{lm}(\hat{r},\hat{r}') &=& \sum_{m_1,m_2} \lla l_1 m_1 l_2 m_2 | l m \rra
Y_{l_1}^{m_1}(\hat{r}) Y_{l_2}^{m_2}(\hat{r}') \cr
Y_{l_1}^{m_1}(\hat{r}) Y_{l_2}^{m_2}(\hat{r}') &=& \sum_{l=|l_1-l_2|}^{l_1+l_2} \sum_{m=-l}^{l}
\lla l_1 m_1 l_2 m_2 | l m \rra \mathcal{Y}^{l_1 l_2}_{lm}(\hat{r},\hat{r}')~,
\end{eqnarray}
and using the Wigner-Eckart theorem, the nonlocal density becomes
\begin{eqnarray}
\label{1.1.4}
\hspace*{-1cm} \rho_{sf}(\vec{r},\vec{r'}) = & &\sum_{l l'} \sum_{K=|l-l'|}^{l+l'} (-1)^{J'-M'}
\threej{J'}{K}{J}{-M'}{k}{M} \mathcal{Y}_{Kk}^{*ll'}(\hat{r},\hat{r}') \cr
& &\times \lla A \lambda' J' M' \left| \left| \sum_{i=1}^A \frac{\delta(r_i-r)}{r^2}
\frac{\delta(r'_i-r')}{r'^2} \mathcal{Y}_{K}^{ll'}(\hat{r}_i,\hat{r}_i') \right| \right| A \lambda J M \rra~.
\end{eqnarray}
We can immediately make a simplification since  in M-scheme calculations $M'=M$. 
Thus, the condition $-M'+k+M=0$ in the $3j$-symbol forces $k$ to be zero.

To further evaluate the nonlocal density, we rewrite Eq.~(\ref{1.1.4}) in second quantization form using
$\alpha$ and $\beta$ as final and initial single-particle HO states, denoted by the single-particle quantum numbers
($n, l, j, t_z$). Then $(a^{\dagger}_{\alpha} \tilde{a}_{\beta})^{(K)}$, where $a_{nljmt_z}=(-1)^{j-m}\tilde a_{n l j -m t_z}$, represents the single-particle transition operator of rank $K$. 
Using the general  expression of the matrix elements of a one-body operator 
$T_K=\sum_i T_{K,i}$ of rank $K$~\cite{Suhonen},
\begin{equation}
\label{1.1.5}
\lla \psi_f; J_f\left|\left| T_K \right|\right| \psi_i; J_i \rra = \frac{1}{\hat{K}} \sum_{\alpha\beta} \lla \alpha \left|\left| T_{K,1} \right|\right| \beta \rra \lla \psi_f; J_f \left|\left| (a^{\dagger}_{\alpha} \tilde{a}_{\beta})^{(K)} \right|\right| \psi_i; J_i \rra~,
\end{equation}
with $\hat{K} = \sqrt{2K+1}$ and $T_{K,1}$ being a single-particle operator,  
 we obtain for the nonlocal density,
\begin{eqnarray}
\label{1.1.6}
\hspace*{-1cm} \rho_{sf}(\vec{r},\vec{r'}) &=& \sum_{l l'} \sum_{K=|l-l'|}^{l+l'} (-1)^{J'-M}
\threej{J'}{K}{J}{-M}{0}{M} \mathcal{Y}_{K0}^{*l l'}(\hat{r},\hat{r}') \times \\
& &\frac{1}{\hat{K}} \sum_{\alpha\beta} \lla \alpha \left|\left| \frac{\delta(r_1-r)}{r^2}
\frac{\delta(r'_1-r')}{r'^{2}} \mathcal{Y}_{K}^{l l'}(\hat{r}_1,\hat{r}'_1) \right|\right| \beta \rra \lla A \lambda' J' \left|\left| (a^{\dagger}_{\alpha} \tilde{a}_{\beta})^{(K)} \right|\right| A \lambda J \rra~. \nonumber
\end{eqnarray}
In Eq.~(\ref{1.1.6}), $\lla A \lambda' J' \left|\left| (a^{\dagger}_{\alpha} \tilde{a}_{\beta})^{(K)} \right|\right| A \lambda J \rra$ are reduced one-body density matrix (OBDM) elements.
They are calculated using NCSM eigenstates $|A\lambda J M\rangle$ and $|A\lambda'
J'M'\rangle$, and are input to our calculations.
Replacing $\alpha$ and $\beta$  by $(n', l'_\alpha, j')$ and $(n, l_\beta, j)$,
respectively, the reduced single-particle matrix element can be obtained using the HO single-particle wavefunctions. Note that, for simplicity, the isospin projections are dropped from the labels, for which $(t_z)_\alpha=(t_z)_\beta$, with only protons entering into calculations of charge densities, while calculations of matter densities involve a summation over both protons and neutrons.
We can thus separate and define the $K$-tensor dependence by
\begin{eqnarray}
\label{1.1.7}
\hspace*{-2cm} 
{\rho}_{l l'K }(r,r') &\equiv & 
\sum_{n j n' j'} \hat{j}\hat{j}' (-1)^{l'+l+j+\half+K} \sixj{l'}{l}{K}{j}{j'}{\half} R_{n' l'}(r') R_{n l}(r) \lla A \lambda' J' \left|\left| (a^{\dagger}_{n' l' j'} \tilde{a}_{n l j})^{(K)} \right|\right| A \lambda J \rra,
\end{eqnarray}
where $R_{nl}(r)$ is the radial component of the single-particle harmonic oscillator wave function (defined in Appendix~\ref{appendixA}). Using Eq.~(\ref{1.1.7}), the matrix elements of $\rho_{sf}(\vec{r},\vec{r'})$ can be expressed
 as a sum over all tensors ${\rho}_{l l' K}(r,r')$,
\begin{equation}
\label{1.1.8}
\rho_{sf}(\vec{r},\vec{r'}) = \sum_{Kll'} (-1)^{J'-M} \threej{J'}{K}{J}{-M}{0}{M}
\mathcal{Y}_{K0}^{*l'l}(\hat{r},\hat{r}') {\rho}_{ll'K}(r,r'),
\end{equation}
separating out the radial and angular components of the nonlocal density.


\subsubsection{Space-fixed nonlocal one-body density matrix in momentum space}

In order to remove the c.m. contribution, we need a momentum space representation of the
nonlocal density, $\rho_{sf}(\vec p, \vec {p'})$. 
We obtain it by applying a  Fourier transformation to $\rho_{sf}(\vec r, \vec {r'})$,
\begin{eqnarray}
\label{1.1.9}
{\rho}_{sf}(\vec{p},\vec{p'}) &=& \frac{1}{(2\pi)^3} \int \int {\rho}_{sf}(\vec{r},\vec{r'}) e^{i\vec{p} \cdot \vec{r}} e^{-i\vec{p'} \cdot \vec{r'}} d^3r d^3r'~,
\end{eqnarray}
where a normalization factor  $\sqrt{\frac{1}{(2\pi)^3}}$  is included for each integral, and 
\begin{equation}
\label{1.1.10}
e^{-i \vec{p} \cdot \vec{r}} = 4\pi \sum_{Cc} Y^{c}_{C}(\hat{r}) Y^{*c}_{C}(\hat{p}) (-i)^{C}
j_C(pr).
\end{equation}
Using  $\rho_{sf}(\vec r, {\vec r}~')$ from Eqs.~(\ref{1.1.8}) and~(\ref{1.1.7}), 
and the orthonormality of the spherical harmonics,
$\int Y_C^c (\hat r) Y_l^{*m} (\hat r) d \hat r= \delta_{lC}\delta_{mc}$~,
leads to 
\begin{eqnarray}
\label{1.1.15}
\rho_{sf}(\vec{p},\vec{p'}) &=& \sum_{Kll'} (-1)^{J'-M} \threej{J'}{K}{J}{-M}{0}{M} \mathcal{Y}_{K0}^{*l'l}(\hat{p},\hat{p}') {\rho}_{l l' K}(p,p'),
\end{eqnarray}
where the radial and angular part of the  $ {\rho}_{l l' K}(p,p')$ is given by
\begin{eqnarray}
\label{1.1.16}
	\rho_{l l' K}(p,p') = \sum_{n j n' j'} \hat{j}\hat{j}' (-1)^{\frac{l-l'}{2}} (-1)^{j+\half} \sixj{l'}{l}{K}{j}{j'}{\half} R_{n' l'}(p') R_{n l}(p) \lla A \lambda' J' \left|\left| (a^{\dagger}_{n' l' j'} \tilde{a}_{n l j})^{(K)} \right|\right| A \lambda J \rra~.
\end{eqnarray}
This expression is used to calculate the densities from  NCSM calculations of OBDM
directly in momentum space. For completeness, the derivation of the local density directly in momentum space is given in Appendix~\ref{appendixA}.

\subsection{Translationally Invariant Nonlocal Densities}
\label{tiformal}

In order to analyze the charge and mass distribution inside the nucleus and employ 
the nonlocal density, e.g., in reaction calculations, it needs to be
translationally invariant. 
In NCSM calculations in a HO basis with  $N_{\max}$ truncation, as well as in the SA-NCSM,
the wave function in single particle coordinates exactly factorizes in a c.m.
wave function and a $ti$ wave function,
\begin{equation}
\label{2.1.1}
 \left| \Psi J M \rra = \left| \Psi_{ti} J M \rra \otimes \left|
\phi_{c.m.} 0s \rra,
\end{equation}
which can be used to remove the c.m. contribution from the nonlocal density.
If we want to extend the scheme for removing the c.m. contribution developed for local
densities~\cite{Navratil:2004dp,Cockrell:2012vd,Dytrych:2015yxa} to the nonlocal case, we need to carefully
consider in which variables we want to work. While in Section~\ref{sfixed}, the nonlocal density
is calculated as a function of the independent momenta $\vec{p}$ and $\vec{p}~'$, 
it is more convenient to proceed with  the independent momenta
\begin{eqnarray}
\label{2.1.2}
       \vec{q} &=& \vec{p~}'-\vec{p} \cr
       \vec{\mathcal{K}} &=& \half (\vec{p~}' + \vec{p})~.
\end{eqnarray}
The corresponding set of coordinate space variables is given by
\begin{eqnarray}
\label{2.1.3}
	\vec{\zeta} &=& \half(\vec{r}+\vec{r~}') \cr
	\vec{Z} &=& \vec{r~}'-\vec{r}~,
\end{eqnarray}
where the displacement $\vec Z$ is translationally invariant. Thus the c.m. contribution must 
only be associated with $\vec{\zeta}$,
\begin{eqnarray}
\label{2.1.4}
       \zeta &=& \zeta_{rel} + \zeta_{c.m.} .
\end{eqnarray}
Because of the exact factorization
of the c.m. wave function and the $ti$ wave function, the $sf$ density
can be expressed as a convolution of the $ti$ density distribution
$\rho_{ti}$ with the c.m. density distribution $\rho_{c.m.}$ via:
\begin{equation}
 \rho_{sf}(\vec{\zeta},\vec{Z}) = \int \rho_{ti}(\vec{\zeta} -
\vec{\zeta}_{c.m.},\vec{Z}) \; \rho(\vec{\zeta}_{c.m.},0) \; d^3\zeta_{c.m.}.
 \end{equation}
 
Based on the set of variables  from Eqs.~(\ref{2.1.2}) and (\ref{2.1.3}), we use a
Fourier transformation of the operator defined in Eq.~(\ref{1.1.1}) and the coordinates defined in
Eq.~(\ref{2.1.2}),
\begin{eqnarray}
\label{2.1.5}
	 \rho_{sf}(\vec{q},\vec{\mathcal{K}}) &=& \frac{1}{ (2\pi)^3}\lla \Psi' J' M \left| \sum_{i=1}^A e^{-i\vec{q} \cdot (\vec{\zeta}_{rel,i} + \vec{\zeta}_{c.m.})} e^{-i\vec{\mathcal{K}} \cdot \vec{Z}_i} \right| \Psi J M \rra \cr
&=& \frac{1}{ (2\pi)^3}\lla \Psi' J' M \left| e^{-i\vec{q} \cdot \vec{\zeta}_{c.m.}} \sum_i e^{-i\vec{q} \cdot \vec{\zeta}_{rel,i}} e^{-i\vec{\mathcal{K}} \cdot \vec{Z}_i} \right| \Psi J M \rra~.
\end{eqnarray}
In the above derivation we employed for the c.m. coordinate
${\vec R}_{c.m.}=\frac{1}{A}\sum_i {\vec r}_i$ and defined 
${\vec \zeta}_{c.m.}=\frac{1}{2}({\vec R}_{c.m.} + {\vec R}'_{c.m.})$. 
Using Eq.~(\ref{2.1.1}) we can separate the c.m. contribution from the intrinsic part
of the nonlocal density.
\begin{flalign}
\label{2.1.6}
	\rho_{sf}(\vec{q},\vec{\mathcal{K}})  = \lla \phi_{cm} 0s | e^{-i\vec{q} \cdot \vec{\zeta}_{c.m.}} | \phi_{cm} 0s \rra
	\frac{1}{ (2\pi)^3} \lla \Psi'_{ti} J' M \left| \sum_i e^{-i\vec{q} \cdot \vec{\zeta}_{rel,i}}
e^{-i\vec{\mathcal{K}} \cdot \vec{Z}_i} \right| \Psi_{ti} J M \rra~. \\ \nonumber
	&
\end{flalign}
We now can define the $ti$ matrix elements for the nonlocal density as
\begin{eqnarray}
\label{2.1.7}
 \rho_{ti}(\vec{q},\vec{\mathcal{K}}) 
\equiv 
\frac{1}{ (2\pi)^3} \lla \Psi'_{ti} J' M \left| \sum_i e^{-i\vec{q} \cdot \vec{\zeta}_{rel,i}} e^{-i\vec{\mathcal{K}} \cdot
\vec{Z}_i} \right| \Psi_{ti} J M \rra.
\end{eqnarray}
Thus, if we know the space-fixed nonlocal density as a function of the momenta $\vec{q}$ and
$\vec{\mathcal{K}}$ and calculate the c.m. contribution in the $|0s\rangle$ state, we
obtain the translationally invariant density. 

Let us first consider the calculation of $\rho_{sf}(\vec{q},\vec{\mathcal{K}})$.
In order to transform the $sf$ nonlocal density to the coordinates $\vec{q}$ and
$\vec{\mathcal{K}}$, 
the harmonic oscillator lengths must be transformed to  $b_{\vec{q}}$ and $b_{\vec{\mathcal{K}}}$. 
This  transformation is explicitly given in Appendix~\ref{appendixB}.
Then we need to
express the product $R_{nl} (p) R_{n'l'}(p') \mathcal{Y}(\hat{p},\hat{p}')$ from Eq.~(\ref{1.1.15}) as a function of $\vec{q}$ and $\vec{\mathcal{K}}$. To do so, we use  Talmi-Moshinsky transformations 
from $\left| n l n' l' : K \rra$ to $\left|n_\mathcal{K}, l_\mathcal{K}, n_q, l_q : K \rra$. Those
Talmi-Moshinsky brackets only depend on the transformation parameter $d$, defined in
Appendix~\ref{appendixB}, the multipole $K$, and the harmonic oscillator quantum numbers ($n,l,n',l'$). They do not depend on $M$ and require the energy conservation $2n'+l'+2n+l = 2n_{\mathcal{K}}+l_{\mathcal{K}}+2n_q+l_q$. 
Thus the
radial  and angular components of the wave functions transform as
\begin{eqnarray}
\label{2.1.8}
\lefteqn {R_{n'l'}(p') R_{nl}(p) \mathcal{Y}_{KM}^{l'l}(\hat{p},\hat{p}') =} &&  \cr
 && \sum_{n_q,n_{\mathcal{K}},l_q,l_{\mathcal{K}}} \lla n_{\mathcal{K}} l_{\mathcal{K}}, n_q l_q : K | n' l', n l : K \rra_{d=1} R_{n_{\mathcal{K}} l_{\mathcal{K}}}(\mathcal{K}) R_{n_q l_q}(q) \mathcal{Y}_{KM}^{l_{\mathcal{K}} l_q}(\hat q,\hat{\mathcal{K}})~.
\end{eqnarray}
With this, the $sf$ nonlocal density as a function of $\vec{q}$ and $\vec{\mathcal{K}}$ becomes
\begin{eqnarray}
\label{2.1.9}
	\hspace*{-2cm} \rho_{sf}(\vec{q},\vec{\mathcal{K}}) &=& \sum_K \sum_{n_q l_q n_{\mathcal{K}} l_{\mathcal{K}}} \sum_{nln'l'jj'} \lla n_q l_q n_{\mathcal{K}} l_{\mathcal{K}} : K | n' l' n l : K \rra_{d=1} (-1)^{J'-M} \threej{J'}{K}{J}{-M}{0}{M} \mathcal{Y}_{K0}^{l_q l_{\mathcal{K}}}(\hat{q},\hat{\mathcal{K}}) \cr 
	& &(-1)^{\frac{l-l'}{2}}  (-1)^{j+\half} \hat{j} \hat{j}'
\sixj{l'}{l}{K}{j}{j'}{\half} R_{n_q l_q}(q) R_{n_{\mathcal{K}}
l_{\mathcal{K}}}(\mathcal{K}) \lla A \lambda' J' \left| \left| (a^{\dagger}_{n'l'j'}
\tilde{a}_{nlj})^{K} \right| \right| A \lambda J \rra \cr
	&=& \sum_K \sum_{l_q l_{\mathcal{K}}} (-1)^{J'-M} \threej{J'}{K}{J}{M}{0}{M}
\mathcal{Y}_{K0}^{l_q l_{\mathcal{K}}}(\hat{q},\hat{\mathcal{K}}) \rho_{l_q l_{\mathcal{K}}
K}(q,\mathcal{K}),
\end{eqnarray}
where the  $K$-tensor component that depends on $q$ and $\mathcal{K}$ is given by
\begin{eqnarray}
\label{2.1.10}
	\rho_{l_q l_{\mathcal{K}} K}(q,\mathcal{K}) &\equiv & \sum_{n_q n_{\mathcal{K}}} \sum_{nln'l'jj'} \lla n_q l_q n_{\mathcal{K}} l_{\mathcal{K}} : K | n' l' n l : K \rra_{d=1} (-1)^{\frac{l-l'}{2}}  (-1)^{j+\half} \cr
	& & \hat{j} \hat{j}' \sixj{l'}{l}{K}{j}{j'}{\half} R_{n_q l_q}(q)
R_{n_{\mathcal{K}} l_{\mathcal{K}}}(\mathcal{K}) \lla A \lambda' J' \left| \left|
(a^{\dagger}_{n'l'j'} \tilde{a}_{nlj})^{K} \right| \right| A \lambda J \rra~.
\end{eqnarray}

\noindent
Next, we calculate the contribution of the c.m. as
\begin{equation}
\label{cmcontrib}
\lla \phi_{c.m.} 0s | e^{-i\vec{q} \cdot \vec{\zeta}_{c.m.}} | \phi_{c.m.} 0s \rra =
 e^{-\frac{1}{4A} b^2 q^2},
\end{equation}
where $A$ is the number of nucleons and $b$ the harmonic oscillator length. The explicit calculation is
given in Appendix~\ref{appendixC}.

Collecting the information from Eqs.~(\ref{2.1.9}) and (\ref{cmcontrib}), the $ti$ nonlocal 
density can be calculated as
\begin{eqnarray}
\label{tiOBDM}
\rho(\vec{q},\vec{\mathcal{K}}) &=& e^{\frac{1}{4A} b^2 q^2} \; \rho_{sf}(\vec{q},\vec{\mathcal{K}})~,
\end{eqnarray}
where we dropped the subscript $ti$.
Since the c.m. contribution is a simple analytic function of $q^2$, the numerical effort in computing the
$ti$ nonlocal density is the computation of the $sf$ nonlocal density
as a function of $\vec{q}$ and $\vec{\mathcal{K}}$ in Eq.~(\ref{2.1.8}).   This is an
important advantage of the current method, and avoids the need for transforming NCSM's OBDM elements to
relative (Jacobi) coordinates.
Subsequently, we can obtain the
$ti$ nonlocal density in coordinate space by a Fourier transformation of Eq.~(\ref{tiOBDM}).   
The $ti$ local density can be computed from Eq.~(\ref{tiOBDM}) by integrating
the nonlocal density over $\vec{\mathcal{K}}$,
\begin{equation}
  \label{tilocal}
  \rho_{K=0}(q) = \int d\mathcal{K} \mathcal{K}^2 \rho_{000}(q,\mathcal{K})~.
\end{equation}
Note that this local density in momentum space is also referred to as the elastic form factor (see e.g. Ref. \cite{Dytrych:2015yxa}), and can also be obtained as the Fourier transform of the local probability density in coordinate space.  We use this as numerical check; in particular, the value at $q=0$ should correspond to the number of nucleons.


\section{Results and Discussion}
\label{results}

\subsection{Nonlocal Densities in Momentum Space}
\label{nlmomentum}

The nonlocal densities shown in this work are based on {\it ab initio}
NCSM or SA-NCSM calculations of OBDM elements that employ the JISP16 $NN$
interaction~\cite{Shirokov:2004ff}.
Before elaborating on the nonlocal structure of the translationally
invariant  density, we first
establish that its construction is consistent with a translationally invariant
local density directly constructed in momentum space as outlined in
Appendix~\ref{appendixA}. 
 In Fig.~\ref{fig1} we show the $K=0$ component
of the local  
proton density  of $^{12}$C as a function of the momentum
transfer $q$, which is constructed as outlined in Appendix~\ref{appendixA}. We also
confirmed that this is numerically equivalent to the local density construction
presented in Ref.~\cite{Cockrell:2012vd}. The solid line represents the Fourier transform of the density to momentum space, the formfactor, which is normalized at $q=0$ to the number of
protons. The solid triangles represent the same quantity obtained by
integrating the nonlocal density  over the nonlocal variable $\mathcal{K}$ according
to Eq.~(\ref{tilocal}). The integrated
values agree with the directly constructed ones at least within six significant
figures.
For comparison, we also include a local density
obtained from a
more traditional Hartree-Fock-Bogolyubov mean-field calculation which
utilizes the density-dependent finite-range {\it Gogny D1S} 
nucleon-nucleon interaction~\cite{Berger:1989bdp,Gogny}. Based on this density 
elastic proton scattering off $^{12}$C was successfully
calculated in~\cite{Chinn:1995qn}. However, a slight mismatch in the diffraction minima 
of the differential cross section could indicate that the slower fall-off of the NCSM
local density may be preferable.

Next, we want to study nonlocal one-body densities of four different
nuclei, the open shell nuclei  $^6$Li and $^{12}$C, and the closed shell nuclei  $^4$He and  $^{16}$O. 
It is well known that $^{12}$C consisting of six
protons and six neutrons is deformed in its body-fixed frame, $^6$Li consisting of three protons and three
neutrons can be sometimes viewed as consisting of a $^4$He core and an additional
 neutron-proton pair in the $p$-shell, while $^4$He and $^{16}$O are closed shell nuclei. In a shell-model framework, the protons
and neutrons in $^4$He occupy predominantly the $s$-shell, while in $^{16}$O they occupy predominantly the $s$- and $p$-shells.
Thus we want to explore if nonlocal
properties reflect some of the common perceptions about those nuclei. Since we make a multipole
expansion of the nonlocal density, it is convenient to concentrate on a specific multipole. 
Here we chose the $K=0$ multipole, since this component determines the density of the $0^+$ ground states for the even-even nuclei under consideration and dominates in the $1^+$ ground state in  $^6$Li.
First, we want to 
consider the  $K=0$ multipole (see Eq.~(\ref{2.1.10}) for notation) of the 
nonlocal one-body density  in momentum space, 
$\rho_{l_q l_\mathcal{K} (K=0)} (q,\mathcal{K})$. 
We show proton densities in physically relevant variables, the momentum 
transfer $q=|{\vec p}~'- \vec{p}|$ and 
$\mathcal{K}=\frac{1}{2}|\vec{p}~'+\vec{p}|$. Note that the variable $\mathcal{\vec K}$, being the conjugate
coordinate to the nonlocal coordinate $\vec Z$, only appears when nonlocal densities are considered. 
The converged nonlocal densities
$\rho_{l_q  l_\mathcal{K} (K=0)} (q,\mathcal{K})$ 
for the proton
distributions of $^{12}$C, $^{16}$O, and $^4$He are displayed  as function of $q$ and $\mathcal{K}$ in Figs.~\ref{fig2},
\ref{fig3}, and \ref{fig4}, respectively, while the corresponding nonlocal density of $^6$Li is given
in the last column of Fig.~\ref{fig7}.
To illustrate the contributions of different angular momenta we show slices
of constant values of $l_q = l_{\mathcal{K}}$. The constraints given
in Eq. (\ref{2.1.8}) indicate that, for $K=0$, once $l_q$ is fixed, $l_{\mathcal{K}}$ takes the
same value. We also need to point out that the contributions of odd values
of $l_q$  cancel out exactly for $K=0$ as a result of the symmetry properties of the Talmi-Moshinsky brackets
\cite{Kamuntavicius:2001pf,moshinsky:1996:harmonic}. 
The vanishing contribution for odd $l_q$ is validated numerically as well.

A common observations for all four nuclei is that the contribution of $l_q=0$ dominates.  
We further  point out that when
integrating  $\rho_{l_q  l_\mathcal{K} (K=0)} (q,\mathcal{K})$ over $\mathcal{K}$ 
to obtain the local density, the
constraints for $l_q$ determine that only $l_q=0$ contributes to the local
density as it is  e.g. given for $^{12}$C in Fig.~\ref{fig1}. Thus, all higher values of $l_q$ 
are genuinely nonlocal contributions for $K=0$.  
For both, $^6$Li and $^4$He, the maximum of the $l_q=0$ slice is located at $q=0$ and
$\mathcal{K}=0$, and the functions fall off smoothly  to zero in $q$ as well  $\mathcal{K}$.
This is different for  $^{12}$C and $^{16}$O, for which the maximum value is located 
around $\mathcal{K}\sim 1$~fm$^{-1}$ and $q=0$,
and which, in addition, exhibit a minimum around $q\sim 2$~fm$^{-1}$ for  $\mathcal{K}=0$.

Figs.~\ref{fig2}-\ref{fig4} as well as Figs.~\ref{fig7}-\ref{fig9} are plotted in a way 
that $\rho_{l_q l_\mathcal{K}}(q=2\mathcal{K},\mathcal{K})$ is shown along
the diagonal. For the closed shell nucleus $^4$He (Fig.~\ref{fig4}), for which the protons are assumed to
dominantly occupy the $s$-shell,  the maximum of the nonlocality for $l_q\ge 2$
follows the diagonal line and moves to higher values of $q$ the larger $l_q$ becomes. For $^6$Li
(Fig.~\ref{fig7}, 4th column), which can
still be considered as dominated by particles in the $s$-shell, the maximum of $\rho_{l_q l_\mathcal{K}}(q, \mathcal{K})$ 
moves slightly
away from the `diagonal' and the off-shell structure can be roughly located between  $\rho_{l_q
l_\mathcal{K}}(q=4\mathcal{K},\mathcal{K})$ and 
$\rho_{l_q l_\mathcal{K}}(q=\mathcal{K},\mathcal{K})$, while for $^{12}$C (Fig.~\ref{fig2}) and
 $^{16}$O (Fig.~\ref{fig3}) the entire nonlocality
is located in this `wedge'. Furthermore the density changes sign along the line $q=2\mathcal{K}$. 
This pattern, together with the different $l_q =0$ behavior, appears to be a signature for nonlocal densities
in $p$-shell dominated nuclei.

\subsection{Nonlocal Densities in Coordinate Space}
\label{nlcoordinate}

Once we obtain the $ti$ nonlocal density  in momentum
space, we can numerically Fourier transform them to coordinate space. The conjugate 
coordinate to the momentum transfer $q$ is the local coordinate $\zeta
=\frac{1}{2}|\vec{r}'+\vec{r}|$, and to $\mathcal{K}$ the nonlocal coordinate
$Z=|\vec{r}'-\vec{r}|$. 
The angular
momenta related to $\zeta$ and $Z$ are denoted as $l_\zeta$ and $l_Z$.
In Fig.~\ref{fig5} we show angular momentum slices of  nonlocal proton densities in coordinate space
for $K=0$
as a function of $\zeta$ and $Z$ for $^{12}$C, and  in Fig.~\ref{fig6} for $^6$Li.  Here we choose
the local coordinate $\zeta$ as one of the axes, so that one can directly read off
the local densities of $^{12}$C and $^6$Li  along the line
$Z=0$.  Hence, Fig.~\ref{fig5} shows that the local density of $^{12}$C has its
maximum at $\zeta \sim 1$ fm and is suppressed at $\zeta =0$ fm, suggesting that the density is pushed away from
the center; indeed, if one plots this density in a body-fixed frame, it will have a deformed  torus-like
shape with a suppressed density in the center. However, the present densities are not calculated in a body-fixed
frame, and  Fig.~\ref{fig5} does not reveal any features that can be associated with the nuclear deformation. On the other hand, we
can say, that when we compare the $l_q=0$ contributions in momentum space of  $^{12}$C and $^{16}$O,
the local density of $^{16}$O  also has a maximum that is  pushed away from the origin, which is a consequence of the $p$-shell being filled up.

The range of the $\zeta$ and $Z$ axes are chosen such that the diagonal of the plot shows
 $\rho_{l_\zeta l_Z}(\zeta,Z=2\zeta)$. Both figures show similar behavior for $l_\zeta \ge 2$ as the
corresponding $q$-$\mathcal{K}$-figure. In Fig.~\ref{fig5} for $l_\zeta \ge 2$, the nonlocal
density changes sign along $\rho_{l_\zeta l_Z}(\zeta,Z=2\zeta)$, and the maxima/minima are roughly located 
in the area given by the lines $\zeta=4Z$ and $\zeta=Z$, indicating a possible $p$-shell dominance of the
nonlocal density. For $^6$Li the situation is slightly different: for $l_\zeta =2$ the maximum
still follows the diagonal and only for $l_\zeta \ge4$ a $p$-shell dominance develops. This may indicate
that the lower $l_\zeta$ are still $s$-shell dominated, while the $p$-shell proton mainly gives the
$l_\zeta=6$ contribution. 

A further study of
NCSM calculations with different $NN$ interactions will have to be carried out to investigate
if the observed nonlocal structures persist and are essentially an indication of the shell structure 
of the nucleus under consideration.

\subsection{Dependence of the Nonlocal Density on the Model Space}
\label{convergence}

The calculations presented in the previous sub-sections are carried out with one-body density
matrix elements from NCSM and SA-NCSM calculations that are close to convergence with  respect to the ground
state binding energies  and low-lying excited states, as far as the model space is concerned.  For $^6$Li those studies
are carried out in Ref.~\cite{Cockrell:2012vd,Dytrych:2015yxa,Dytrych:2016cc}, together with model-dependence  studies with
respect to the root-mean-square point-proton radius and the quadrupole moment.  The calculations for $^{12}$C are
discussed in Ref.~\cite{Dytrych:2016cc} and those for $^{16}$O in
Refs.~\cite{Maris:2013poa,Maris:2008ax}.
The $ti$ density should become independent of the
basis parameters $\hbar\omega$ and $N_{\max}$ as $N_{\max}$ increases.
However, it is impractical to carry out a convergence study of the
nonlocal density itself, and instead, we illustrate how some of the
features of the nonlocal density depend on the model space.

First, we consider the $K=0$ component of the nonlocal proton density for a
fixed $\hbar\omega$~=~20~MeV as a function of the $N_{\rm max}$ truncation.
In Fig.~\ref{fig7} we display slices of $\rho_{l_q l_\mathcal{K}} (q,\mathcal{K})$ for
$^6$Li for fixed
$l_q=l_\mathcal{K}$ for $N_{\rm max}$~=~6, 10, 12, and 14.
For $l_q$~=~0 we observe
that the density maximum at $q=0$, $\mathcal{K}=0$ increases as a function of $N_{\rm max}$.
This is consistent with the fact that the tail of the wavefunction in coordinate space
(and hence radii) become better described as $N_{\rm max}$ increases.  However,
the general distribution remains the same.  The angular momentum slices $l_q$~=~2 and $l_q$~=~4
clearly show how the nonlocal structure builds up as $N_{\rm max}$ increases, but again,
the general distribution remains the same.  
Even going from
$N_{\rm max}$~=~12 to $N_{\rm max}$~=~14, there is a very slight increase of the maxima of $\rho_{l_q l_\mathcal{K}}$.
The slice for $l_q$~=~6 for $N_{\rm max}$~=~6 exhibits a very different nonlocal
structure in comparison to the higher $N_{\rm max}$ values, which can be understood as 
an effect of the model-space truncation. 
We observe changes in the nonlocal structure for $l_q$~=~6 even
when going from $N_{\rm max}$~=~12 to $N_{\rm max}$~=~14.  However,
the absolute values of this contribution is so small, that
calculations of observables based on this nonlocal density are
unlikely to be affected by the $l_q$~=~6 (or higher) contributions.
In general, the figure shows that the $N_{\rm max}$~=~6 model space is
not sufficient to describe nonlocal correlations, but the nonlocal
structure looks reasonably well converged at $N_{\rm max}$~=~14,
similarly to the results calculated earlier for the binding energy and
other observables~\cite{Cockrell:2012vd}.

Next we keep $N_{\rm max}$ fixed ($N_{\rm max}$=12) and study the
nonlocal structure as a function of the oscillator parameter
$\hbar\omega$, where we choose the values 15~MeV, 20~MeV, and 25~MeV.
It is well known that, for a given $N_{\rm max}$, the basis truncation introduces effective infrared (IR) cutoff and ultraviolet (UV) cutoffs that also depend on the $\hbar\omega$ value: namely, very low $\hbar\omega$ values (a small momentum UV cutoff) cut out high momenta that may affect short-range correlations, while very large $\hbar\omega$ values (small spatial IR cutoff) affect the wavefunction tail~\cite{Coon:2012ab,More:2013rma,Wendt:2015nba,FurnstahlHP12}. Note that increasing $N_{\rm max}$ increases both the IR and UV cutoffs, removing both cutoffs in the infinite model space limit. Exploring the $\hbar\omega$ dependence, in comparison to the large $N_{\rm max}$ limit of Fig.~\ref{fig7} which improves both cutoffs, can provide some indication if the nonlocality is sensitive to these cutoffs.
The corresponding functions $\rho_{l_q l_\mathcal{K}} (q,\mathcal{K})$ for
$l_q=l_\mathcal{K}$~=~0, 2, 4, and 6 are shown in Fig.~\ref{fig8} and
$\rho_{l_\zeta l_Z}(\zeta,Z)$ in Fig.~\ref{fig9} for the proton
density of $^6$Li.  The study of Ref.~\cite{Cockrell:2012vd} has
already shown that for the local density $\rho(r)$ of $^6$Li, a
smaller value of $\hbar\omega$ leads to a better description of the
asymptotic tail of the wavefunction and a spatially expanded density,
but the density in the nucleus interior becomes low, whereas for
larger values of $\hbar\omega$ the situation is reversed.  Indeed,
the larger $\hbar\omega$ value yields a significantly lower maxima
for the density in momentum space (Fig.~\ref{fig8}), whereas the smaller
$\hbar\omega$ value gives very pronounced maxima as function of $q$
and $\mathcal{K}$.  As already mentioned above, this is the result of
the poor convergence of the tail of the wavefunction in coordinate
space for large values of $\hbar\omega$.  On the other hand, low
$\hbar\omega$ values lack a good description of the high momentum
behavior.  Furthermore, for low $\hbar\omega$
the maxima are moved toward low $q$ momentum transfers, and for high
$l_q$ a particle is mainly transferred from (or to) low $p$ momentum
(diagonal line, as discussed above).

Considering the coordinate space density, $\rho_{l_\zeta l_Z}(\zeta,Z)$,
in Fig.~\ref{fig9}, we find that for small $\hbar\omega$ values, the
nonlocal structures are well developed at larger values of $\zeta$ and
$l_\zeta$, while for $l_\zeta$~=~0 the maximum at $\zeta=Z=0$ is less
developed.  Again, this is consistent with the findings in
Ref.~\cite{Cockrell:2012vd}.  In contrast, for $\hbar\omega$~=~25~MeV,
we notice that more features are resolved in the nonlocal structure,
especially for higher $l_\zeta$, a direct result of improving the UV
cutoff.  Overall, we conclude that the nonlocality of the density seems
to be more sensitive to the IR cutoff, that is, to the description of
the tail of the wavefunction in coordinate space, except for high
$l_q$, where the nonlocal contribution, while being very small,
becomes also sensitive to the UV cutoff.

\subsection{Study of the Nonlocality of the Density}
\label{nlspecific}

To study the nonlocal  behavior in a more 
quantitative fashion, we plot 
the $K=0$, $l_q=0$  component of $\rho_{l_q l_\mathcal{K}} (q,\mathcal{K})$
for fixed values of $q$ as a
function of $\mathcal{K}$ for $^{12}$C (Fig.~\ref{fig10} (a)) and $^4$He ((Fig.~\ref{fig10} (b)). 
As soon as we take $\mathcal{K}$-slices of $^{12}$C at higher values  of 
$q$, the form of the nonlocality changes, dips for $q=1$~fm$^{-1}$ for small $\mathcal{K}$ and becomes
negative for even larger $q$. 
Since the magnitude of $\rho_{K=0,l_q=0}(q,\mathcal{K})$  changes by orders of magnitude when
moving along $q$, we normalize the 
slices by a factor $N$ given by
\begin{equation}
  N = \frac{\rho(q=0,\mathcal{K}=0)}{\rho(q,\mathcal{K}=0)}.
\label{normalize}
\end{equation}
We also
note, that $\rho_{K=0,l_q=0}(q,\mathcal{K})$  falls off quickly as a function of $\mathcal{K}$, 
independent of the
value of $q$ and becomes essentially zero for $\mathcal{K} \geq 2$~fm$^{-1}$.  Comparing with 
Fig.~\ref{fig3}, the nonlocal density of $^{16}$O exhibits the same behavior. 
Panel (b) of Fig.~\ref{fig10} shows  
similar slices of the $K=0$ component of $\rho_{l_q l_\mathcal{K}} (q,\mathcal{K})$  for
$^4$He. Here we find that the nonlocal density is positive for all values of $q$ and
falls off like a Gaussian. However, there is no uniform Gaussian bell shape
for all $q$, since for the larger $q$ values, the Gaussian width increases. It appears that there is
no simple parameterization of this behavior as a function of $q$.
Similarly to $^{12}$C, the nonlocality of the $^4$He density  is essentially zero
for $\mathcal{K} >2$~fm$^{-1}$, though it has larger high-momentum components compared to 
$^{12}$C. This can be understood from realizing that smaller radii in coordinate
space translate to larger high-momentum components.

Finally, we show in Fig.~\ref{fig11}(a) a `formfactor'  $\rho_{K=0} (\mathcal{K})$, for $K=0$, as a 
function of $\mathcal{K}$, where 
$\rho_{l_q l_\mathcal{K}} (q,\mathcal{K})$ is integrated over $q$. It is worthwhile noting that for
$^4$He and $^6$Li this function is positive, while it starts as negative values for $^{12}$C and
$^{16}$O before turning positive. This, together with the observations of Sec.~\ref{nlmomentum}, may allow one to
conclude that if a nucleus is dominated by $s$-shell nucleons, the value of $\rho_{K=0} (\mathcal{K}=0)$
is positive, and when $p$-shell nucleons dominate, $\rho_{K=0} (\mathcal{K}=0)$ is negative. 

In addition, we show in 
Fig.~\ref{fig11}(b) the conventional proton formfactors (local densities in momentum space) for the same nuclei, which are
normalized to the proton number at $q=0$. Only the charge distributions of the heavier nuclei have a
zero crossing visible in the figure, the one for $^6$Li turns negative at $\sim 6$~fm$^{-1}$ while it
stays completely positive for $^4$He. Generally, the proton formfactor provides information about the
spatial charge distribution of the nucleus. The information given by $\rho_{K=0} (\mathcal{K})$ 
gives a consistent picture, namely, after the $s$-shell is filled, additional protons fill up the $p$-shell.

\section{Conclusions and Outlook}
\label{summary}

In this work we explored features of translationally invariant nonlocal one-body densities obtained
from {\it ab initio} NCSM and SA-NCSM calculations using the JISP16 $NN$ interaction~\cite{Shirokov:2004ff}
 for several light nuclei. In order to do this, we first
defined the nonlocal one-body density in a space-fixed coordinate system in such a way that it
directly relates to the OBDM elements which a NSCM calculation provides, and constructed
space-fixed nonlocal one-body densities for $^4$He, $^6$Li, $^{12}$C, and $^{16}$O in momentum 
and position space. As examples for our study, we chose $^4$He and $^{16}$O  representing closed shell
nuclei, together with  $^6$Li and $^{12}$C  representing open shell nuclei.

To remove the c.m. part of the wavefunctions calculated 
in the NCSM using a harmonic oscillator
basis, we first needed to transform the space-fixed nonlocal densities from conventionally used
linearly independent variables $\vec p$ and $\vec{p}'$ to another linearly independent set $\vec q$
and $\vec{\mathcal{K}}$ which is more appropriate for our task. Their conjugate coordinate variables
$\vec \zeta$ and $\vec Z$ are such that the c.m. contribution is only contained in $\vec \zeta$. 
With this, we can successfully extend a scheme developed for removing c.m. contributions from local one-body
densities~\cite{RochfordD92,Navratil:2004dp,Cockrell:2012vd,Dytrych:2015yxa} to nonlocal one-body densities. 

We studied the nonlocal structure of the one-body densities as a function of the angular momentum $l_q$
in momentum as well as coordinate space. For all four nuclei the largest contribution to the
nonlocal density comes from the $l_q=0$ part, for which the nonlocality is restricted to about 2~fm. 
The higher angular momenta, though at least two orders of magnitude smaller, 
contribute exclusively to the nonlocal structure. Thus nuclear properties or reactions that are
dominated by these angular momentum contributions will show sensitivity to the nonlocality. 
In addition, we found that the nonlocal structure of the neutron and proton one-body densities does not
show any significant difference for the $N=Z$ nuclei we investigated. 
We also found that the structure of the  nonlocality reflects the shell structure of the nuclei we
considered. Once the $p$ shell becomes dominant, the nonlocality exhibits a specific pattern not visible
in the $s$-shell dominated $^4$He. 
Finally, we investigated if there may be some systematic behavior in the nonlocal structure of the
one-body densities which might be captured in some analytic form. 
While this might be possible for the nonlocal structure of $^4$He, it does not look promising for the
other nuclei we investigated.

We note that the current results are presented for the JISP16 $NN$ interaction, and we have found that, e.g.,
using chiral potentials such as the NNLO$_{\rm opt}$ \cite{Ekstrom13} for $^6$Li does not introduce significant changes into the density outcomes presented here. A further study that adopts different $NN$ interactions will have to be carried out to investigate if the observed nonlocal structures persist and are essentially an indication of the nuclear shell structure. We have also studied the role of nonlocality in densities calculated from the SA-NCSM using selected model spaces,  which yields results that are essentially the same as compared to those obtained in the corresponding complete model spaces. This will allow one to study nonlocal density features in heavier nuclear systems. The outcomes of these studies will be the focus of a following publication.

Summarizing, this work shows how the c.m. contribution can be removed from {\it ab initio} nonlocal one-body
densities using NCSM wavefunctions, in a similar way as is known for local ones. This will open the path for those densities
to be employed, for example, in calculations of nuclear reactions.



\appendix


\section{Derivation of the Space-Fixed Local One-Body Density Constructed in Momentum Space}
\label{appendixA}

To apply the procedure for removing the c.m. contribution from the local density as suggested in
Refs.~\cite{Cockrell:2012vd,Dytrych:2015yxa,FurnstahlHP12}, the space-fixed local density constructed in
coordinate space needs to be Fourier transformed to momentum space. A numerical Fourier transform as
suggested in~\cite{Cockrell:2012vd} will introduce numerical errors specifically at large momenta 
due to the highly oscillatory nature of the transformation. 
Therefore, it is highly desirable to derive a scheme in which the space-fixed local density is
constructed directly in momentum space.
For this, we need the HO wave functions, $R_{nl}(p)$, in momentum space
\begin{eqnarray}
\label{A.1}
R_{n l}(p) &=& (-1)^{n} \left[ \frac{2 (b^2)^{l+3/2} \Gamma (n+1)}{\Gamma (n + l +
\frac{3}{2})} \right]^{\half} p^l e^{-\half p^2 b^2} L_n^{l+\frac12}(p^2 b^2)~,
\end{eqnarray}
with  harmonic oscillator length  $b = \sqrt{\frac{\hbar^2 c^2}{mc^2 \hbar \Omega}}$. The corresponding coordinate space HO wave functions are given as
\begin{eqnarray}
\label{A.2}
	R_{n l}(r) = \sqrt{\frac{2}{\pi}} \int dp p^2 R_{n l}(p) j_{l}(rp) =  \left[ \frac{2 \Gamma (n+1)}{(b^2)^{l+3/2} \Gamma (n + l +
\frac{3}{2})} \right]^{\half} r^l e^{-\half \frac{r^2}{b^2}} L_n^{l+\frac12}(\frac{r^2}{b^2})~.
\end{eqnarray}
Here a  normalization coefficient $\sqrt{\frac{2}{\pi}}$ is included.
The function $L_{n}^{l+\half}(\frac{r^2}{b^2})$ represents the associated Laguerre polynomials.
Note the difference in phase of $(-1)^n$ in $R_{nl}(p)$ and $R_{nl}(r)$. 

Combining this with the multipole expansion of the space-fixed nonlocal one-body density in Eq.~(\ref{1.1.6}) and Eq.~(\ref{1.1.7}), we arrive at
\begin{eqnarray}
\label{A.3}
	\rho_{sf}^{(K)}(\vec{r},\vec{r'}) &=& \sum_{nljn'l'j'} \sum_{K=|l-l'|}^{l+l'} (-1)^{J'-M'} \threej{J'}{K}{J}{-M'}{0}{M} \sum_{m,m'} \lla l m l' m' | K 0 \rra Y_l^{*m}(\hat{r}) Y_{l'}^{*m'}(\hat{r}') \times \cr
	& & \hat{j}\hat{j'}\; (-1)^{l'+l+j+\half+K} \sixj{l'}{l}{K}{j}{j'}{\half} \sqrt{\frac{2}{\pi}} \int dp p^2 R_{n l}(p) j_{l}(rp) \sqrt{\frac{2}{\pi}} \int dp' p'^2 R_{n' l'}(p') j_{l}(r'p')  \cr
	& &\lla A \lambda' J' \left|\left| (a^{\dagger}_{n'l'j'} \tilde{a}_{nlj})^{(K)} \right|\right| A \lambda J \rra~.
\end{eqnarray}
Setting  $\vec{r}=\vec{r'}$, reducing the spherical harmonics, and simplifying the resulting
Clebsch-Gordan coefficients by combining them with the 6j symbol leads to
\begin{eqnarray}
\label{A.4}
	\rho_{sf}^{(K)}(\vec{r}) &=& \sum_{nljn'l'j'} \sum_{K=|l-l'|}^{l+l'} (-1)^{J'-M'}
\threej{J'}{K}{J}{-M'}{0}{M} \threej{j'}{j}{K}{\half}{-\half}{0} \frac{1}{\sqrt{4\pi}}
\hat{j}\hat{j'}\; (-1)^{j+\frac{3}{2}+K} Y_K^{*0}(\hat{r}) \cr
	& &  \sqrt{\frac{2}{\pi}} \int dp p^2 R_{n l}(p) j_{l}(rp) \sqrt{\frac{2}{\pi}} \int dp' p'^2 R_{n' l'}(p') j_{l}(rp')  
	 \lla A \lambda' J' \left|\left| (a^{\dagger}_{n'l'j'} \tilde{a}_{nlj})^{(K)} \right|\right| A \lambda J \rra~.
\end{eqnarray}
Rearranging the integrals and performing the Fourier transformation leads to
\begin{flalign}
\label{A.5}
	& \rho_{sf}^{(K)}(\vec{q}) = \sum_{nljn'l'j'} \sum_{K=|l-l'|}^{l+l'} (-1)^{J'-M'}
\threej{J'}{K}{J}{-M'}{0}{M} \threej{j'}{j}{K}{\half}{-\half}{0} \frac{1}{\sqrt{4\pi}}
\hat{j}\hat{j'}\; (-1)^{j+\frac{3}{2}+K}  (i)^{K} \cr
	& Y_K^{*0}(\hat{r}) \lla A \lambda' J' \left|\left| (a^{\dagger}_{n'l'j'}
\tilde{a}_{nlj})^{(K)} \right|\right| A \lambda J \rra 8 \int dp p^2 R_{n l}(p) \int dp'
p'^2 R_{n'l'}(p') \int dr r^2 j_K(qr) j_{l}(rp) j_{l'}(r'p')~.
\end{flalign}
For the special case of $K=0$ the integral over $r$ can be evaluated analytically, noting that $l=l'$ and $j=j'$,
\begin{eqnarray}
\label{A.6}
	\int dr r^2 j_0(qr) j_{l}(rp) j_{l'}(rp') = \frac{\pi}{4} \frac{\beta(\Delta)}{p p' q} P_l(\Delta).
\end{eqnarray}
Here $P_l(\Delta)$ are Legendre polynomials, and the argument $\Delta$ is defined as
\begin{eqnarray}
\label{A.7}
	\Delta = \frac{p^2 + p'^2 - q^2}{2pp'}.
\end{eqnarray}
The function $\beta(\Delta)$ is given as
\begin{eqnarray}
\label{A.8}
	\beta(\Delta)&=&1 ~\text{for} ~-1 < \Delta < 1 \cr
	\beta(\Delta)&=&1/2 ~\text{for}~ \Delta = \pm 1 \cr
	\beta(\Delta)&=&0 ~\text{otherwise}.
\end{eqnarray}
The function $\beta(\Delta)$ allows to constrain  the integral over $p'$ in Eq.~(\ref{A.4}) to the
values
\begin{eqnarray}
\label{A.9}
	p'\leq q+p ~\text{and}~ p' \geq \left| q-p \right| .
\end{eqnarray}
This leads to the final expression for the momentum space local density, which can be calculated
directly in momentum space from given OBDM elements from NCSM calculations,
\begin{flalign}
\label{A.10}
	& \rho_{sf}^{(0)}(q) = \sum_{n n' l j} (-1)^{J'-M'} \threej{J'}{0}{J}{-M'}{0}{M} \threej{j}{j}{0}{\half}{-\half}{0} \sqrt{\pi} \hat{j}\hat{j} (-1)^{j+\frac{3}{2}} Y_0^{*0}(\hat{r}) \cr
	& \lla A \lambda' J' \left|\left| (a^{\dagger}_{n'lj} \tilde{a}_{nlj})^{(0)} \right|\right| A \lambda J \rra \int_0^{\infty} dp p^2 R_{n l}(p) \int_{\left| p-q \right| }^{p+q} dp' p'^2 R_{n' l}(p') \frac{1}{p p' q} P_l(\Delta)~.
\end{flalign}


\section{Derivation of harmonic oscillator lengths for the Transformation to $q$ and $\mathcal{K}$}
\label{appendixB}

For transforming the momenta of
$\rho(\vec{p},\vec{p'})$ in Eq.~(\ref{1.1.15}) to momenta $\vec{q}$ and $\vec{\mathcal{K}}$
we need to know 
how the harmonic oscillator lengths transform. Defining  $b_{\mathcal{K}}$ and $b_{q}$, we can 
infer that a dimensionless coordinate transformation must hold in the same fashion as the coordinate transformation defined in Eq.~(\ref{2.1.2}),
\begin{eqnarray}
\label{B.1}
b_{\mathcal{K}}\vec{\mathcal{K}} &=& \frac{b_{\mathcal{K}}}{2b}b\vec{p}~' +
\frac{b_{\mathcal{K}}}{2b}b\vec{p}~ \cr
b_{q}\vec{q} &=& \frac{b_{q}}{b}b\vec{p}~' - \frac{b_{q}}{b}b\vec{p}.
\end{eqnarray}
The transformation can be written as
\begin{eqnarray}
\label{B.2}
	\begin{pmatrix}
		b_{\mathcal{K}}\vec{\mathcal{K}} \\ b_q \vec{q}
	\end{pmatrix}
	=
	\begin{bmatrix}
		\sqrt{\frac{d}{1+d}} & \sqrt{\frac{1}{1+d}} \\ \sqrt{\frac{1}{1+d}} & -\sqrt{\frac{d}{1+d}}
	\end{bmatrix}
	\begin{pmatrix}
		b\vec{p}~' \\ b\vec{p}
	\end{pmatrix}~,
\end{eqnarray}
with $d$ as a yet undetermined parameter. 
A comparison with Eq.~(\ref{B.1})  leads to
\begin{eqnarray}
\label{B.4}
	\frac{b_{\mathcal{K}}}{2b} &=& \sqrt{\frac{d}{1+d}} = \sqrt{\frac{1}{1+d}} \cr
	\frac{b_q}{b} &=& \sqrt{\frac{1}{1+d}} = \sqrt{\frac{d}{1+d}}~,
\end{eqnarray}
which is then solved as
\begin{eqnarray}
\label{B.5}
	d=1~& ~\rm{and}~  &~b_{\mathcal{K}}=\sqrt{2}b \cr
	d=1~& ~\rm{and}~  &~b_q=\frac{b}{\sqrt{2}}~.
\end{eqnarray}
This transformation of the harmonic oscillator lengths is the same for the conjugate variables  $\vec{\zeta}$ and $\vec{Z}$,
\begin{eqnarray}
\label{B.6}
	d=1~&  ~\rm{and}~  &~b_{Z}=\sqrt{2}b \cr
	d=1~&  ~\rm{and}~  &~b_{\zeta}=\frac{b}{\sqrt{2}}~.
\end{eqnarray}
The values of $d$ enters the  Talmi-Moshinsky brackets in
Eq.~(\ref{2.1.8}), and $b_{q}$ and $b_{\mathcal{K}}$ the radial oscillator functions.


\section{Derivation of the Center of Mass Contribution}
\label{appendixC}

As indicated in  Eq.~(\ref{2.1.4}) the variable  $\zeta$ can be separated 
into a component representing the
relative motion and one for the c.m. motion. The displacement $Z$ is already translationally invariant.
According to Eq.~(\ref{2.1.1}) the c.m. component of (SA-)NCSM eigenstates is exactly
factorized and, by construction, is in
the  $|0s\rangle$ state.
Thus we need to compute Eq.~(\ref{cmcontrib}),
\begin{eqnarray}
\label{C.3}
\lefteqn{\lla \phi_{c.m.} 0s | e^{-i \vec{q} \cdot \vec{\zeta}_{c.m.}} | \phi_{c.m.} 0s \rra } &&\cr
	&=& \int \int d^3 R_{c.m.} d^3 R'_{c.m.} R_{n l}(R_{c.m.}) R_{n' l'}(R'_{c.m.})
\mathcal{Y}_{K0}^{l l'}(\widehat{R_{c.m.}},\widehat{R_{c.m.}}') \;  e^{-i \vec{q} \cdot \vec{\zeta}_{c.m.}} \cr
	&=& \int \int d^3\zeta_{c.m.} d^3Z_{c.m.} \sum_{n_q,n_K,l_q,l_{\mathcal{K}}} \lla
n_\mathcal{K} l_{\mathcal{K}}, n_q l_q : K | n' l', n l : K \rra_{d=1} R_{n_{\zeta} l_{\zeta}}(\zeta_{c.m.}) R_{n_{Z} l_{Z}}(Z_{c.m.}=0) \cr
	& &~~~~~~~~\mathcal{Y}_{K0}^{l_{\zeta} l_{Z}}(\widehat{\zeta_{c.m.}},\widehat{Z_{c.m.}})
 \; e^{-i \vec{q} \cdot \vec{\zeta}_{c.m.}} \cr
&=& \int d^3\zeta_{c.m.} R_{00}(\zeta_{c.m.}) R_{00}(0) \frac{1}{4\pi} \;  e^{-i \vec{q} \cdot 
 \vec{\zeta}_{c.m.}}~.
\end{eqnarray}
We note that
\begin{equation}
\label{C.4}
	\lla n_\mathcal{K}=0~l_{\mathcal{K}}=0, n_q=0~l_q=0 : K=0 | n'=0~l'=0, n=0~l=0 : K=0 \rra_{d=1} = 1.
\end{equation}
Furthermore, if $n=l=n'=l'=0$, then $n_q=l_q=n_{\mathcal{K}}=l_{\mathcal{K}}=0$ as well.
\noindent
Evaluating the radial wave function using Eq.~(\ref{A.2}) with the corresponding harmonic oscillator
lengths, we obtain 
\begin{eqnarray}
\label{C.5}
	R_{0 0}(\zeta_{c.m.}) &=& \left[\frac{2^2}{(b_{\zeta_{c.m.}}^2)^{3/2} \sqrt{\pi}} \right]^{\half} e^{-\half \frac{\zeta_{c.m.}^2}{b_{\zeta_{c.m.}}^2}} \cr
	R_{00}(Z_{c.m.}=0) &=& \left[\frac{2^2}{(b_{Z_{c.m.}}^2)^{3/2} \sqrt{\pi}} \right]^{\half}
\end{eqnarray}
where $b_{\zeta_{c.m.}}^2 = \frac{b_{\zeta}^2}{A}$ and $b_{Z_{c.m.}}^2 = \frac{b_{Z}^2}{A}$.
\noindent
Inserting Eq.~(\ref{C.5}) into Eq.~(\ref{C.3}) leads to
\begin{eqnarray}
\label{C.6}
	\lla \phi_{c.m.} 0s \left| e^{-i \vec{q} \cdot \vec{\zeta}_{c.m.}} \right| \phi_{c.m.} 0s
\rra = \left(\frac{1}{\pi}\right)^{3/2} \frac{1}{(b_{\zeta_{c.m.}})^{3/2}}
\frac{1}{(b_{Z_{c.m.}})^{3/2}} \int d^3\zeta_{c.m.} e^{-\half
\frac{\zeta_{c.m.}^2}{b_{\zeta_{c.m.}}^2} - i\vec{q} \cdot \vec{\zeta}_{c.m.}}~.
\end{eqnarray}
Completing the square in the integral leads to
\begin{eqnarray}
\label{C.7}
\lla \phi_{c.m.} 0s \left| e^{-i \vec{q} \cdot \vec{\zeta}_{c.m.}} \right| \phi_{c.m.} 0s \rra 
 &=& \left(\frac{2 b_{\zeta_{c.m.}}}{b_{Z_{c.m.}}}\right)^{3/2} e^{-\half b_{\zeta_{c.m.}}^2 q^2}
= e^{-\frac{1}{4A} b^2 q^2}~,
\end{eqnarray}
where we used the relations for $b_{\zeta}$ and $b_{Z}$ from Eq.~(\ref{B.6}) to arrive at the final
expression for the c.m. contribution.


\begin{acknowledgments}
We thank J. P. Vary, P. Navr\'{a}til, and T. Dytrych for useful discussions.
This work was performed in part under the auspices of the U.~S.  Department of Energy under contract
Nos. DE-FG02-93ER40756, DE-SC0008485 and DE-SC0018223, 
by the U.S. NSF (ACI-1516338 \& ACI-17136900),
and of DFG and NSFC through funds provided to the
Sino-German CRC 110 ``Symmetries and the Emergence of Structure in QCD" (NSFC
Grant No.~11621131001, DFG Grant No.~TRR110).
A.N. acknowledges support of Ohio University through
the Robert and Ren{\'e} Glidden Visiting Professorship Program,
the Institute of Nuclear and Particle Physics and the Department of Physics and Astronomy. 
Ch.E. acknowledges the hospitality and support of the NSCL at Michigan State
University during the final part of the work. Ch.E., K.L., and G.P. thank the Institute of
Nuclear Theory at the University of Washington for its hospitality during INT-17a, which
stimulated part of this work.
The numerical computations benefited from computing resources provided
by Blue Waters, as well as the Louisiana Optical
Network Initiative and HPC resources provided by  LSU ({\tt www.hpc.lsu.edu}), as well as
resources of the National Energy Research Scientific Computing Center, a DOE
Office of Science User Facility supported by the Office of Science of the U.S. Department of Energy under contract
No. DE-AC02-05CH11231, and 
 JUQUEEN and JURECA of the JSC, J\"ulich, Germany.

\end{acknowledgments}


\bibliography{denspot,clusterpot,ncsm}

\clearpage

\begin{figure}
\centering
\includegraphics[width=10cm]{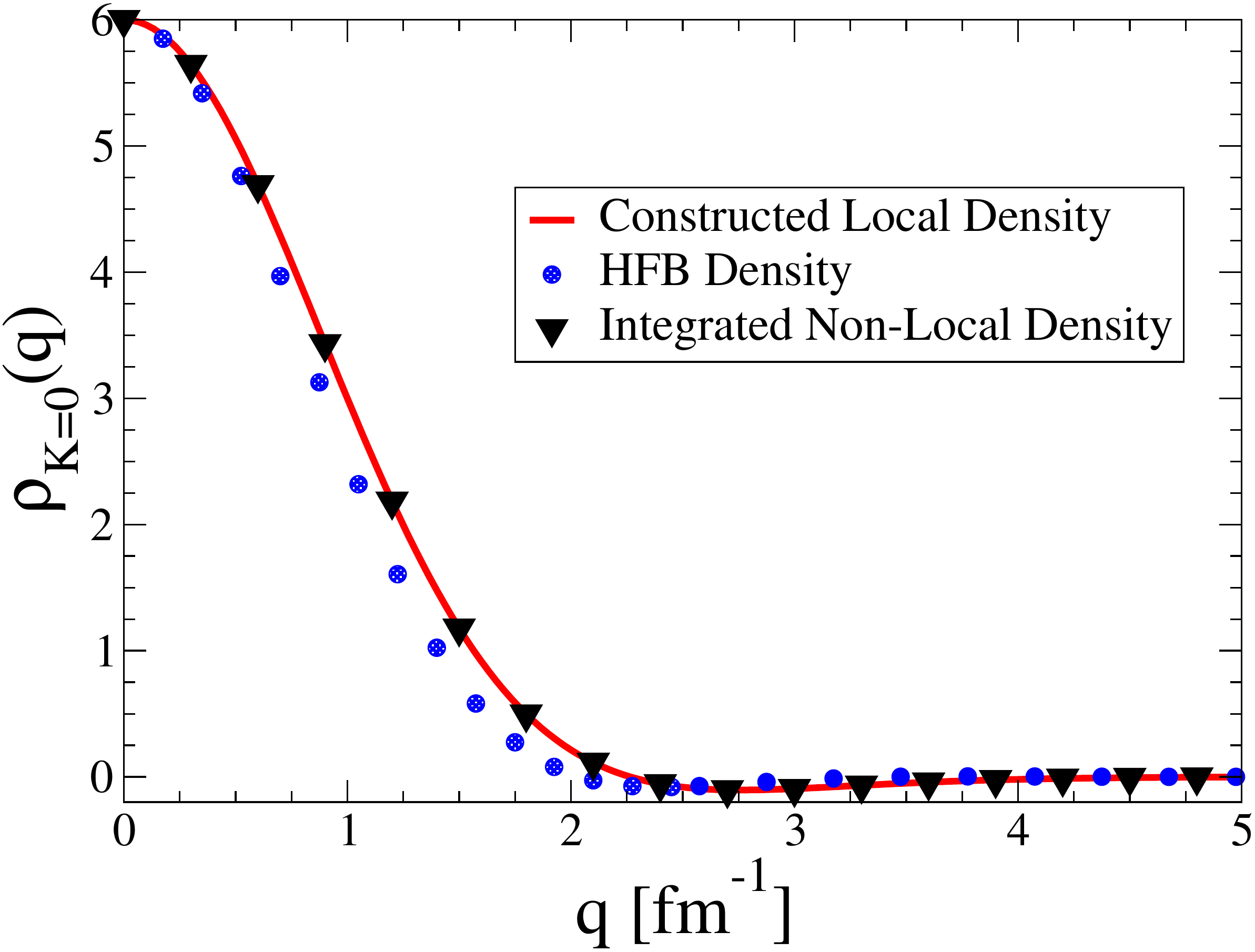}
\caption{The translationally invariant 
local one-body density obtained from a NCSM
calculation ($N_{\max}=10, \hbar\omega=20$~MeV) based on the 
JISP16 $NN$ interaction
for the proton distribution
of $^{12}$C as function of the momentum transfer $q$. The solid line (red)
shows the direct construction in momentum space, while the solid
triangles (black) give the local density obtained by integrating the
nonlocal density over the momentum $\mathcal{K}$. As comparison a local
density obtained from a HFB mean field calculation based on the Gogny
interaction~\cite{Gogny} is shown by the filled solid (blue) circles.
}
\label{fig1}
\end{figure}

\begin{figure}
\centering
\includegraphics[width=14cm]{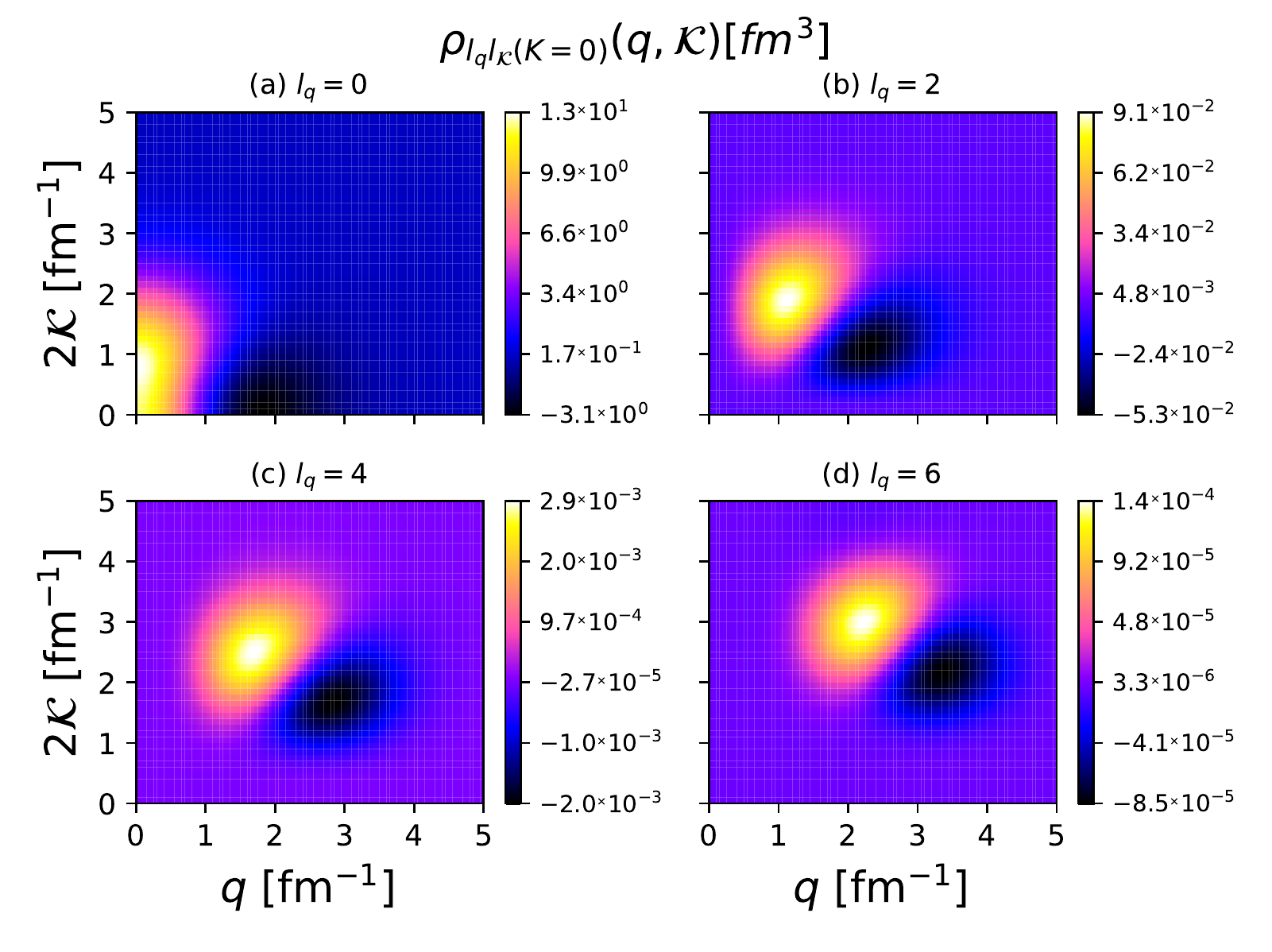}
\caption{The $K=0$ component of the translationally invariant
nonlocal one-body density obtained from a NCSM
calculation ($N_{\max}=10, \hbar\omega=20$ MeV) based on the JISP16 $NN$
interaction
for the proton distribution
of $^{12}$C as function of the momenta $q$ and $\mathcal{K}$.   
Panel (a) depicts the contribution of $l_q=0$, (b) of $l_q=2$, (c) of
$l_q=4$, and (d) of $l_q=6$.
}
\label{fig2}
\end{figure}

\begin{figure}
\centering
\includegraphics[width=14cm]{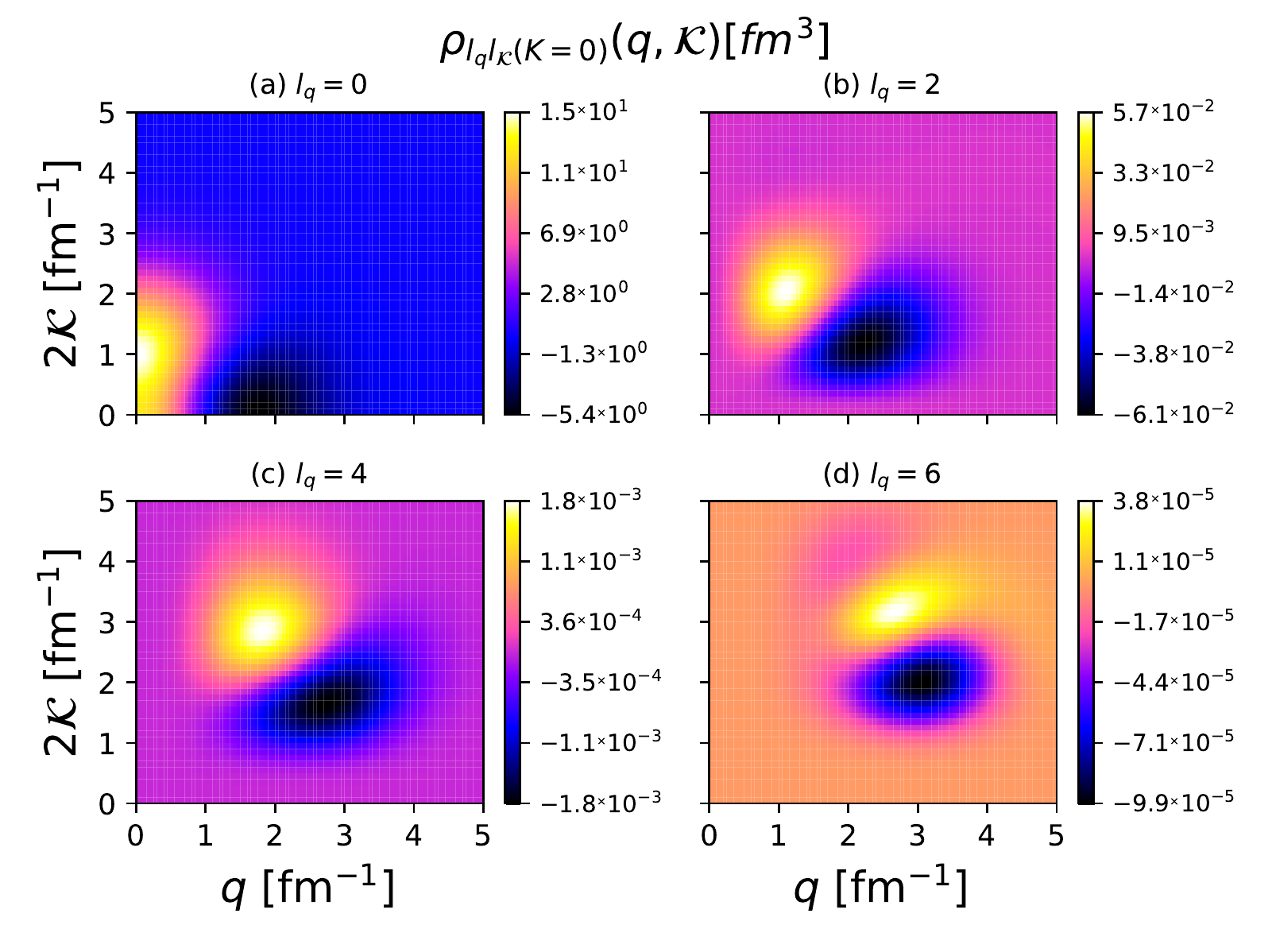}
\caption{
The $K=0$ component of the translationally invariant
nonlocal one-body density obtained from a NCSM 
calculation ($N_{\max}=8, \hbar\omega=20$ MeV) based on the JISP16
$NN$ interaction
for the proton distribution
of $^{16}$O as function of the momenta $q$ and $\mathcal{K}$.   
Panel (a) depicts the contribution of $l_q=0$, (b) of $l_q=2$, (c) of
$l_q=4$, and (d) of $l_q=6$.
}
\label{fig3}
\end{figure}

\begin{figure}
\centering
\includegraphics[width=14cm]{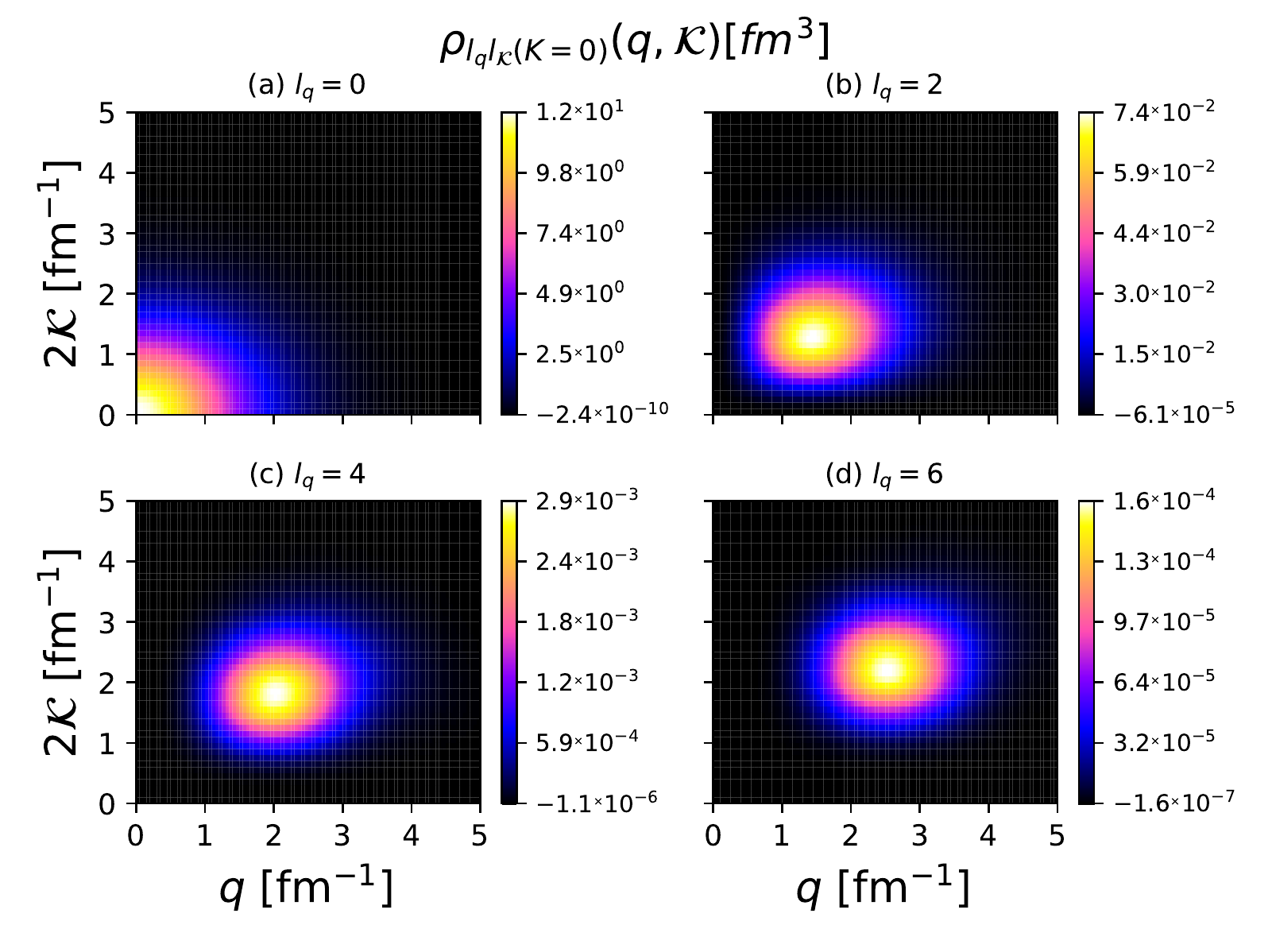}
\caption{
The $K=0$ component of the translationally invariant
nonlocal one-body density obtained from a NCSM
calculation ($N_{\max}=14, \hbar\omega=20$ MeV) based on the JISP16
$NN$ interaction
for the proton distribution
of $^{4}$He as function of the momenta $q$ and $\mathcal{K}$.
Panel (a) depicts the contribution of $l_q=0$, (b) of $l_q=2$, (c) of
$l_q=4$, and (d) of $l_q=6$.
}
\label{fig4}
\end{figure}

\begin{figure}
\centering
\includegraphics[width=14cm]{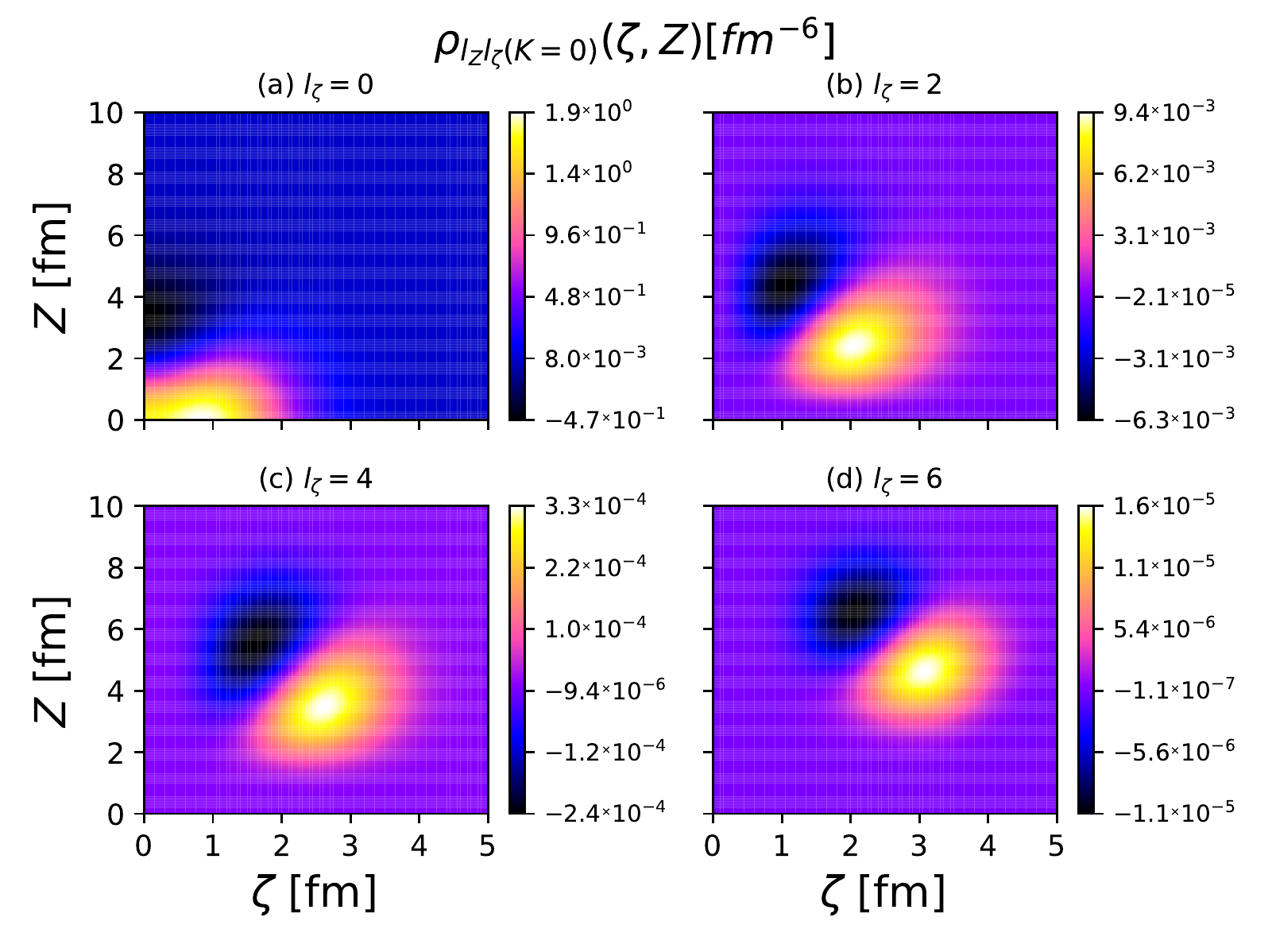}
\caption{The $K=0$ component of the translationally invariant
nonlocal one-body density obtained from a NCSM
calculation ($N_{\max}=10, \hbar\omega=20$ MeV) 
based on the JISP16 $NN$ interaction
for the proton distribution
of $^{12}$C as function of the local coordinate $\zeta$ and the
nonlocal coordinate $Z$. 
Panel (a) depicts the contribution of $l_\zeta=0$, (b) of $l_\zeta=2$, (c) of
$l_\zeta=4$, and (d) of $l_\zeta=6$.
}
\label{fig5}
\end{figure}

\begin{figure}
\centering
\includegraphics[width=14cm]{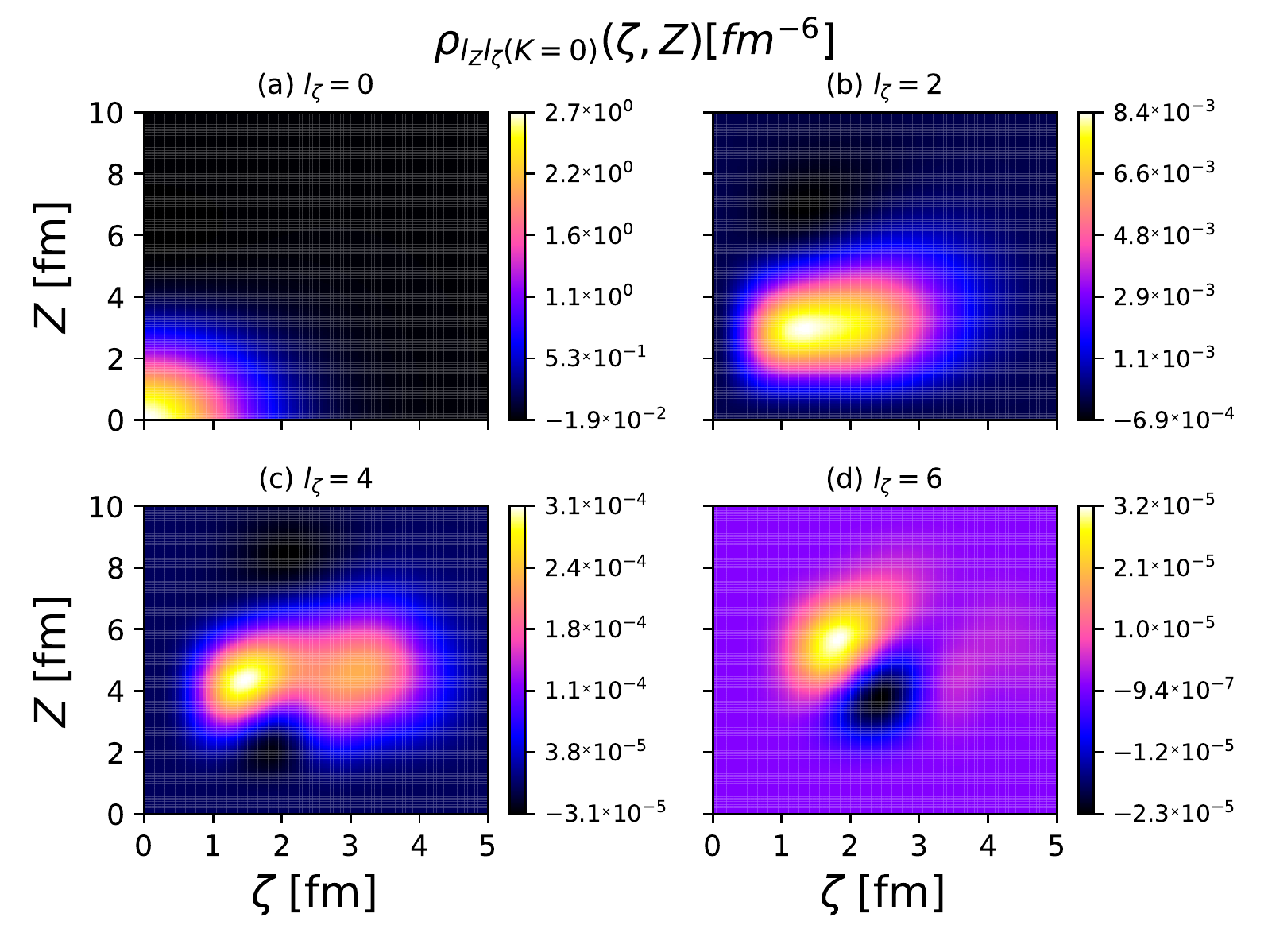}
\caption{
The $K=0$ component of the translationally invariant
nonlocal one-body density obtained from a NCSM
calculation ($N_{\max}=14, \hbar\omega=20$ MeV)
based on the JISP16 $NN$ interaction
for the proton distribution
of $^{6}$Li as function of the local coordinate $\zeta$ and the
nonlocal coordinate $Z$.
Panel (a) depicts the contribution of $l_\zeta=0$, (b) of $l_\zeta=2$, (c) of
$l_\zeta=4$, and (d) of $l_\zeta=6$.
}
\label{fig6}
\end{figure}

\begin{figure}
\centering
\includegraphics[width=17cm]{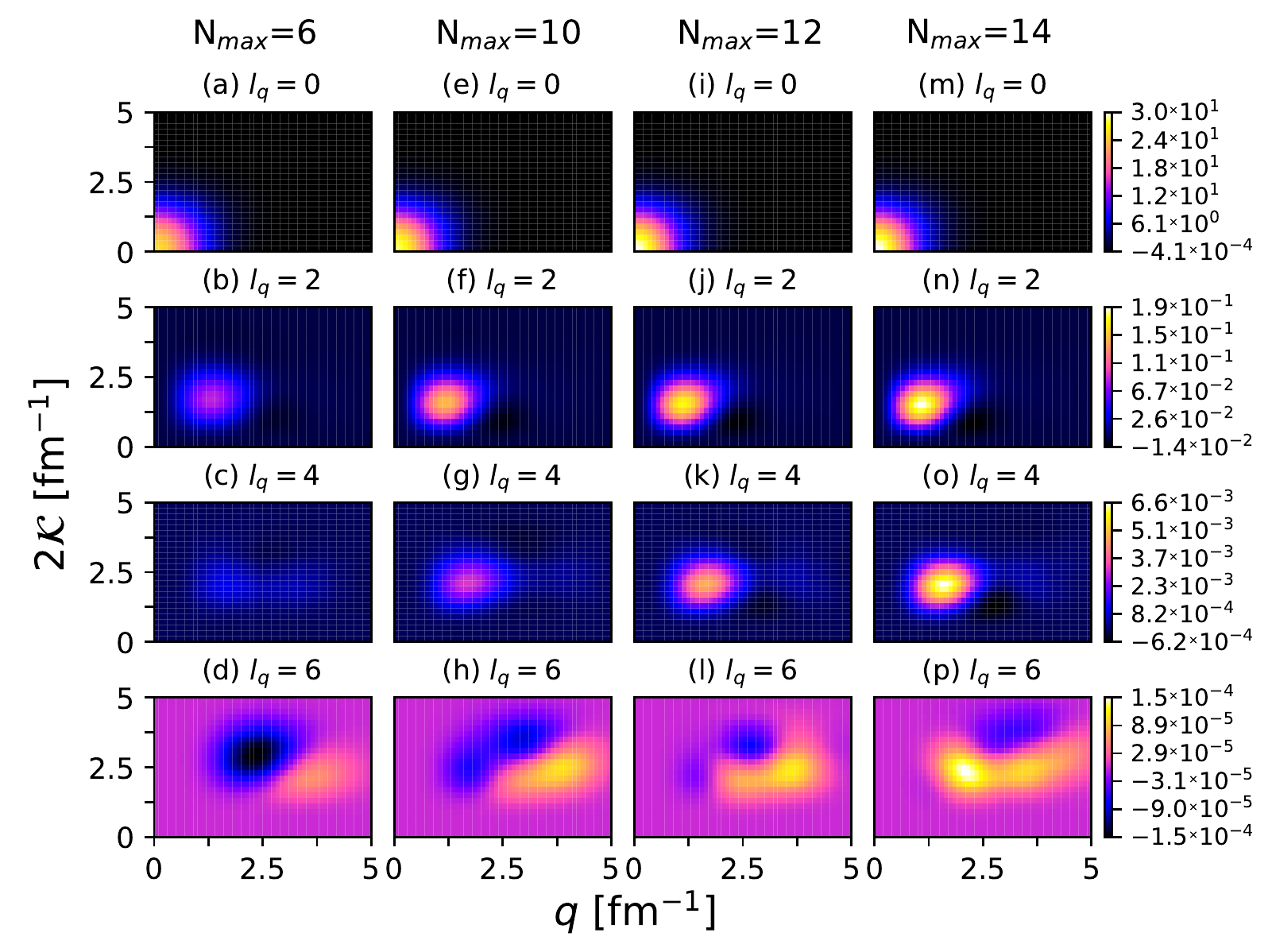}
\caption{
The $K=0$ component of the translationally invariant
nonlocal one-body density $\rho_{l_q l_\mathcal{K}} (q,\mathcal{K})$
obtained from a NCSM
calculation ($\hbar\omega=20$ MeV) based on the JISP16
$NN$ interaction for the proton distribution
of $^{6}$Li as function of the momenta $q$ and $\mathcal{K}$ 
and the size of the model space. The first column contains angular momentum slices
obtained with $N_{\max}=6$, the second with $N_{\max}=10$, the third with $N_{\max}=12$, and
the fourth with $N_{\max}=14$.
The rows represent  different angular momentum slices $l_q=l_\mathcal{K}$ from 0 to 6.
}
\label{fig7}
\end{figure}

\begin{figure}
\includegraphics*[width=17cm]{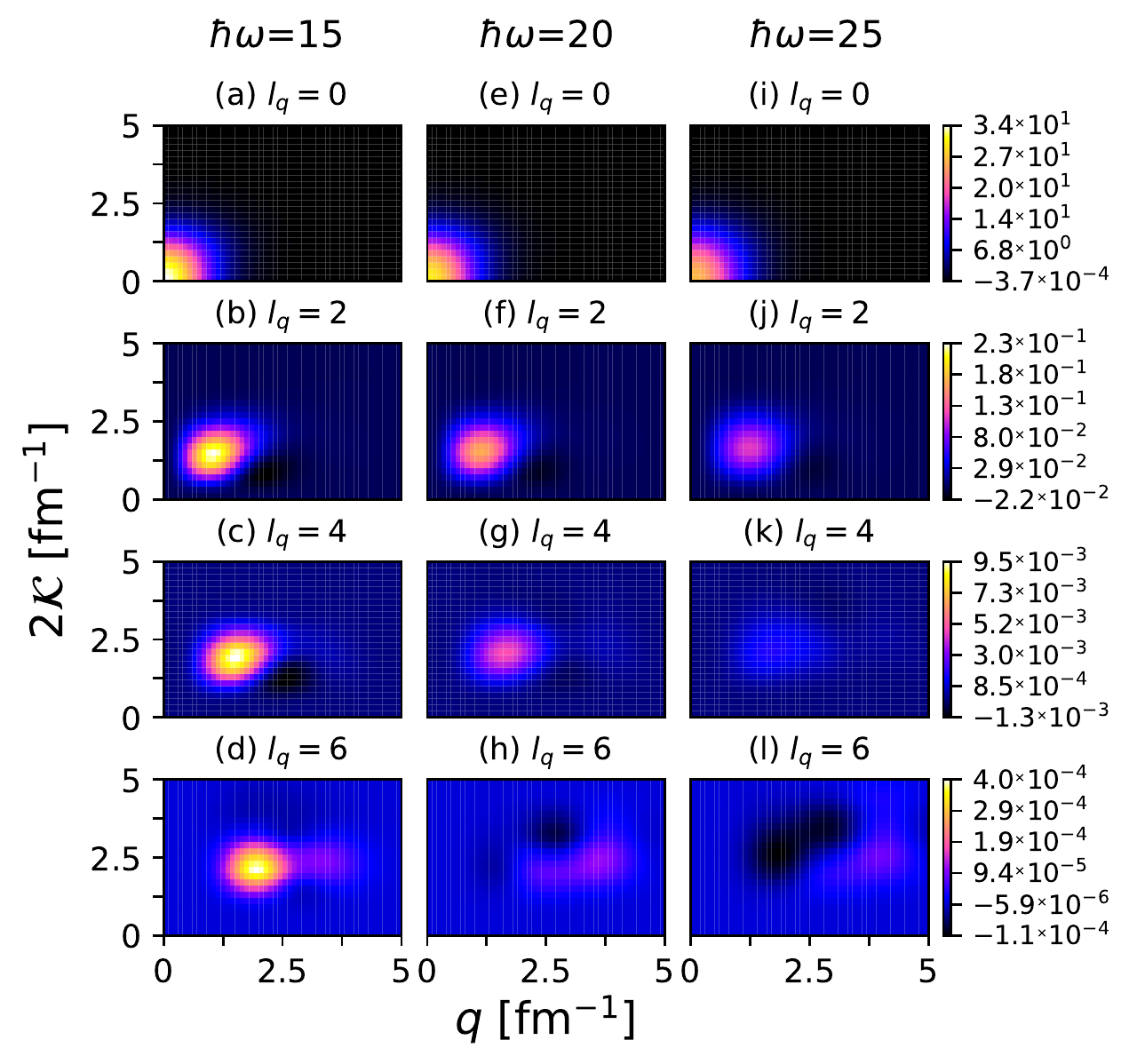}
\caption{
The $K=0$ component of the translationally invariant
nonlocal one-body density $\rho_{l_q l_\mathcal{K}} (q,\mathcal{K})$
obtained from a NCSM
calculation ($N_{\max}=12$) based on the JISP16
$NN$ interaction for the proton distribution
of $^{6}$Li as function of the momenta $q$ and $\mathcal{K}$
and the oscillator parameter $\hbar \omega$. The first column contains angular momentum slices
obtained with $\hbar\omega = 15$~MeV, the second with $\hbar\omega = 20$~MeV, and
the third with $\hbar\omega = 25$~MeV.
The rows represent  different angular momentum slices $l_q=l_\mathcal{K}$ from 0 to 6.
}
\label{fig8}
\end{figure}

\begin{figure}
\includegraphics*[width=17cm]{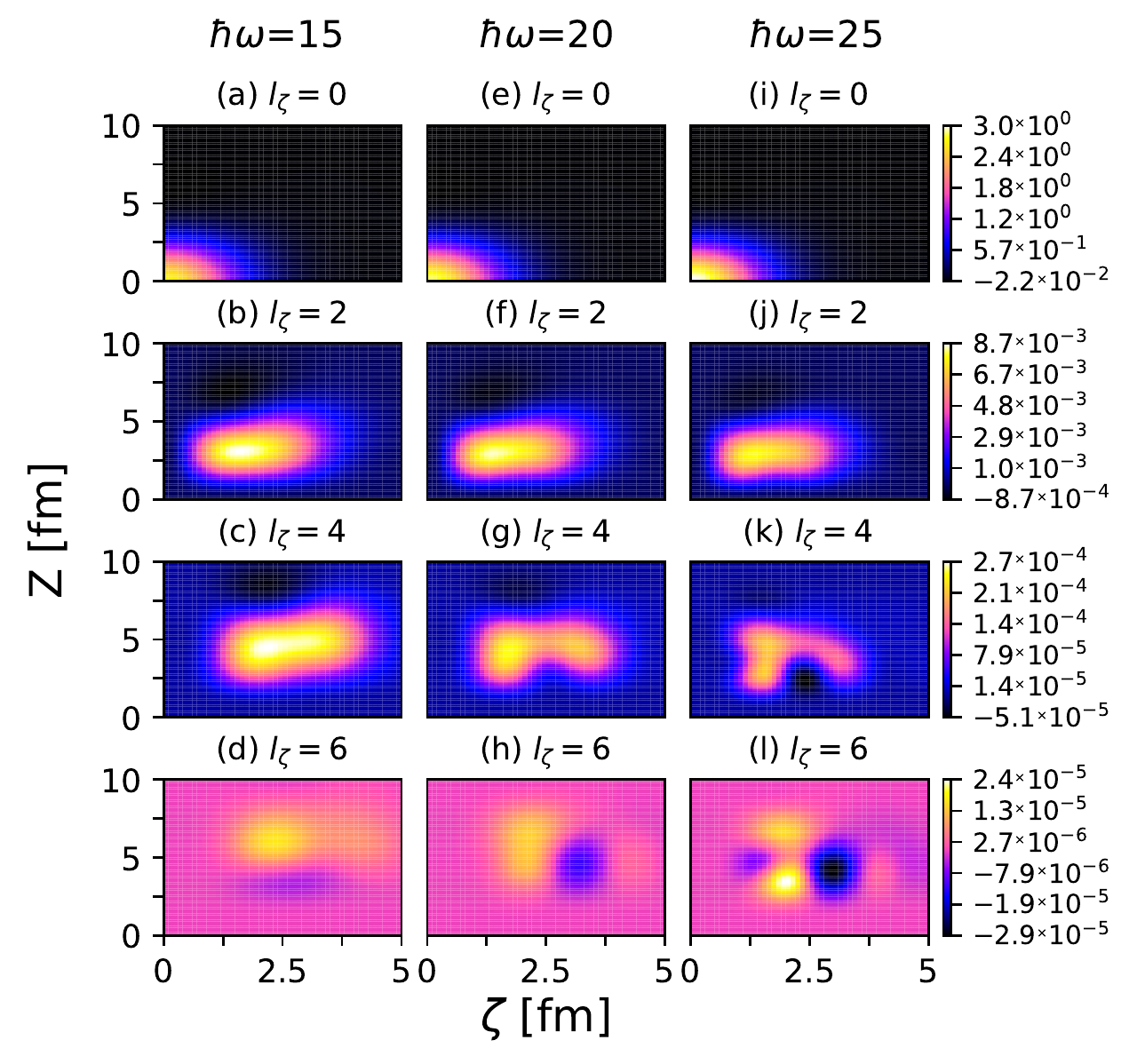}
\caption{ 
The $K=0$ component of the translationally invariant
nonlocal one-body density $\rho_{l_\zeta l_Z} (\zeta,Z)$
obtained from a NCSM
calculation ($N_{\max}=12$) based on the JISP16
$NN$ interaction for the proton distribution
of $^{6}$Li as function of the local coordinate $\zeta$, the nonlocal coordinate $Z$
and the oscillator parameter $\hbar \omega$. The first column contains angular momentum
slices
obtained with $\hbar\omega = 15$~MeV, the second with $\hbar\omega = 20$~MeV, and 
the third with $\hbar\omega = 25$~MeV.
The rows represent  different angular momentum slices $l_\zeta=l_Z$ from 0 to 6.
}
\label{fig9}
\end{figure}

\begin{figure}
\centering
\includegraphics[width=12cm]{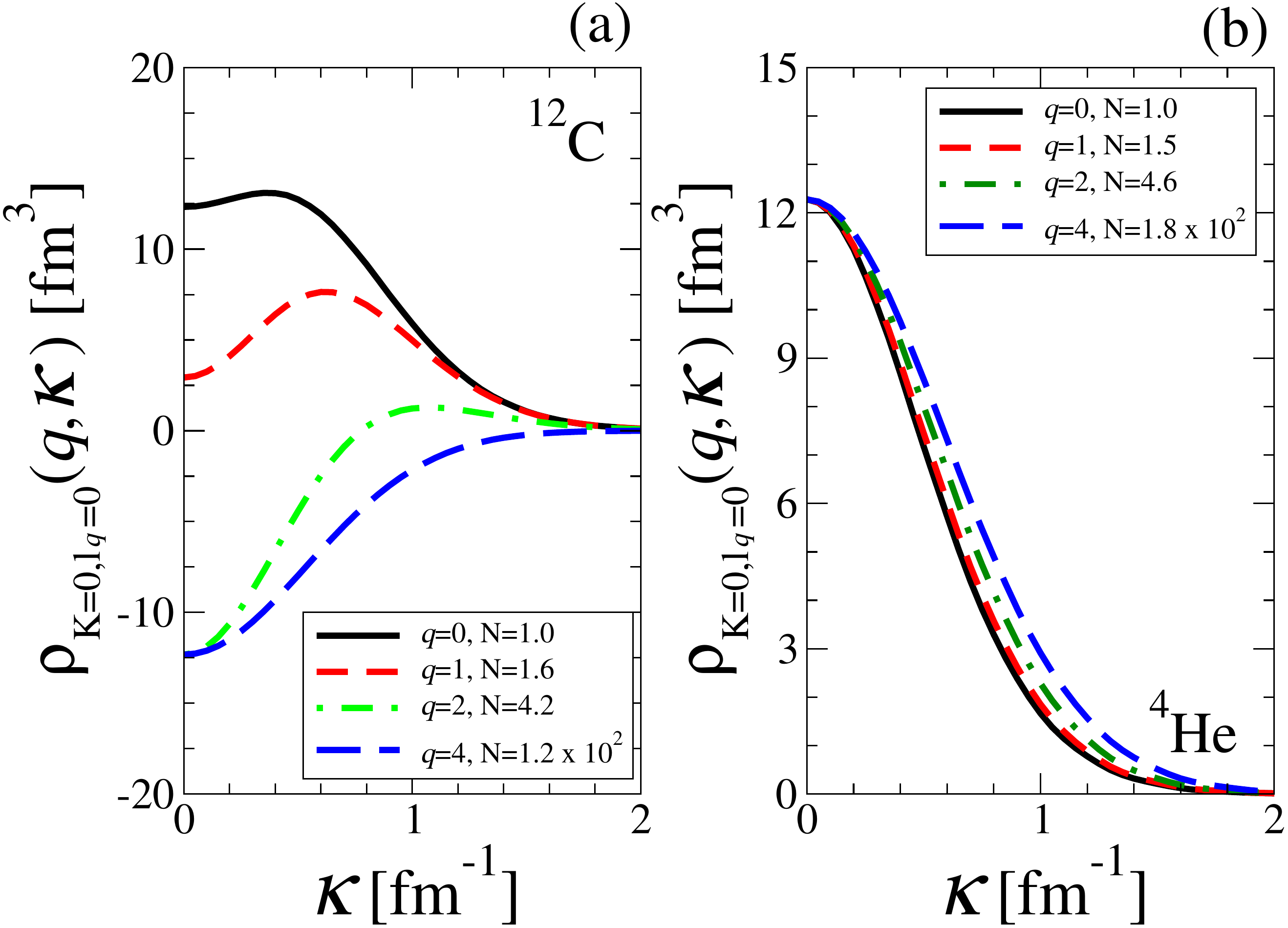}
\caption{The $K=0$, $l_q=0$ component of the translationally invariant 
nonlocal one-body density obtained from a NCSM
calculation 
based on the JISP16 $NN$ interaction
for the proton distribution
of $^{12}$C (panels (a)) and $^{4}$He (panel (b)) as a function of the nonlocal momentum 
$\mathcal{K}$ at fixed momenta $q$ as indicated in the legend.  
The distributions are normalized by the factors indicated in the legend. 
}
\label{fig10}
\end{figure}

\begin{figure}
\centering
\includegraphics[width=12cm]{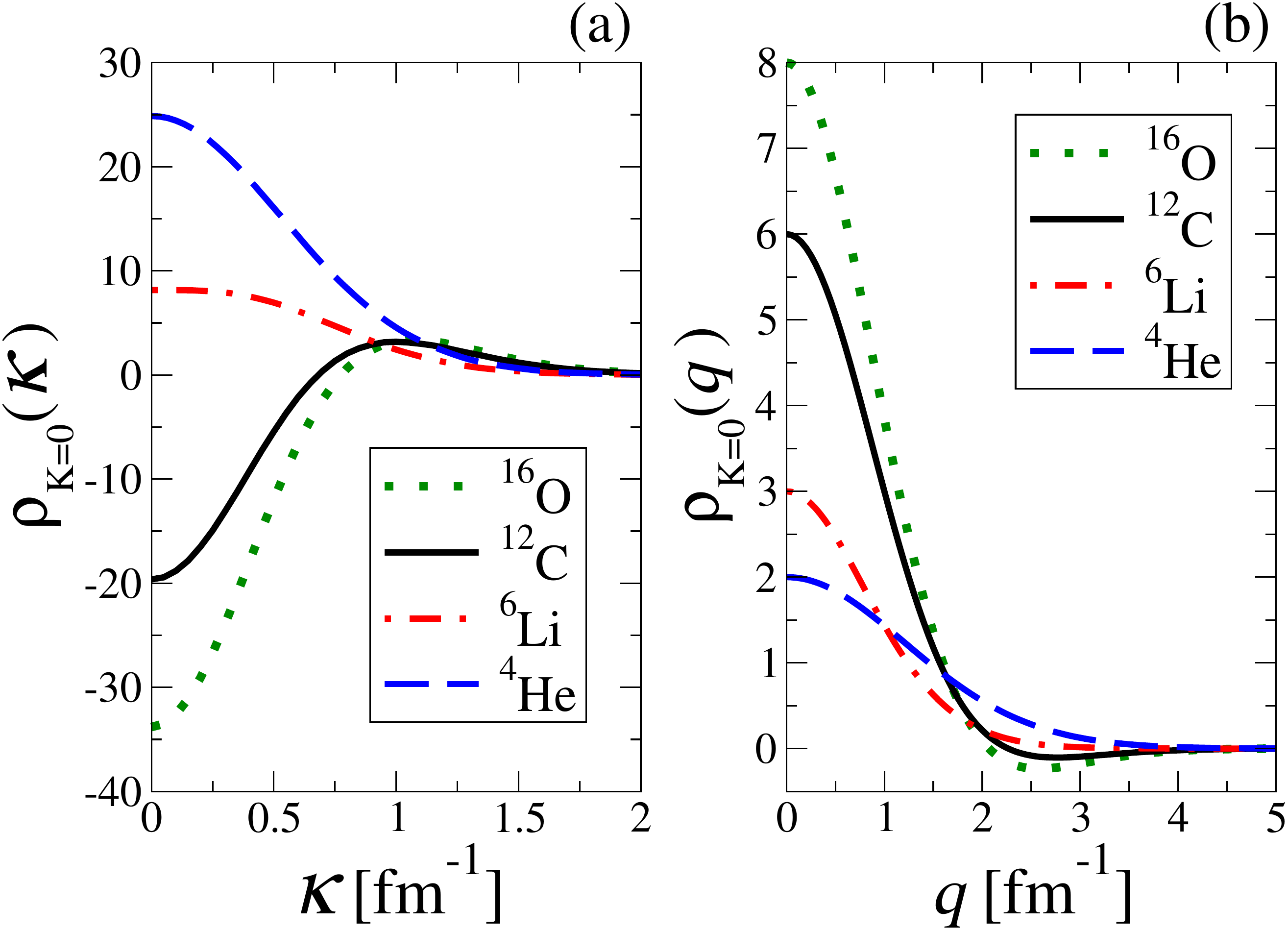}
\caption{The $K=0$ component of the translationally invariant one-body density obtained from  NCSM
calculations based on the 
JISP16 $NN$ interaction
for the proton distributions of $^{4}$He, $^6$Li,  $^{12}$C, and $^{16}$O
 as a function of the nonlocal momentum $\mathcal{K}$ when integrated of the local
momentum $q$ (panel (a)). Panel (b) depicts the
local densities for the same nuclei as function of the momentum transfer $q$ when
integrated over the nonlocal momentum $\mathcal{K}$.
}
\label{fig11}
\end{figure}

\end{document}